\documentclass{article}
\usepackage[utf8]{inputenc}
\usepackage{amsmath}
\DeclareMathOperator*{\argmax}{arg\,max}
\DeclareMathOperator*{\argmin}{arg\,min}
\usepackage{bm}
\usepackage{appendix}
\usepackage{subcaption}
\usepackage{graphicx}
\usepackage{amsfonts}
\usepackage{mathrsfs}
\usepackage{amssymb}
\usepackage{authblk}
\usepackage{booktabs}
\usepackage{titlesec}
\usepackage{geometry}
\usepackage[parfill]{parskip}
\usepackage{scalerel}
\usepackage[natbib=true, style=numeric, sorting=nyt]{biblatex}
\numberwithin{equation}{section}
\newtheorem{theorem}{Theorem}[section]
\newtheorem{lemma}{Lemma}[section]

\newcommand{\bbeta}{\bm{\beta}}
\newcommand{\X}{\bm{X}}
\newcommand{\y}{\bm{y}}
\newcommand{\lnorm}{\|\bm{\beta}^*\|_2}
\newcommand{\tlnorm}{\|\bm{\Tilde{\beta}}\|_2}
\newcommand{\dnorm}{\|\Delta\|_2}
\newcommand{\bb}{\bm{b}}
\newcommand{\bxj}{\bm{x}_j}
\newcommand{\fbxj}{\bm{x}_{1j}}
\newcommand{\sbxj}{\bm{x}_{2j}}
\newcommand{\I}{\mathcal{I}}
\newcommand{\tb}{\Tilde{\bm{\beta}}}
\newcommand{\bs}{\bm{\beta}^*}
\newcommand{\bu}{\bm{u}}
\newcommand{\tu}{\Tilde{\bm{u}}}

\newcommand{\ek}{\bm{e}_k}
\newcommand{\twk}{\Tilde{\bm{w}}_k}
\newcommand{\wk}{\bm{w}_k}
\newcommand{\ftwk}{\Tilde{\bm{w}}_{1k}}
\newcommand{\stwk}{\Tilde{\bm{w}}_{2k}}
\newcommand{\ftwl}{\Tilde{\bm{w}}_{1l}}
\newcommand{\stwl}{\Tilde{\bm{w}}_{2l}}

\title{Constructing Confidence Intervals for the Signals in Sparse Phase Retrieval}
\author{Yisha Yao\thanks{The research of Yisha Yao is partially supported by NSF grants DMS-1721495 and IIS-1741390.}}
\affil{Rutgers University}
\date{}
\begin{document}
\maketitle
\begin{abstract}
In this paper, we provide a general methodology to draw statistical inferences on individual signal coordinates or linear combinations of them in sparse phase retrieval. Given an initial estimator for the targeting parameter (some simple function of the signal), which is generated by some existing algorithm, we can modify it in a way that the modified version is asymptotically normal and unbiased. Then confidence intervals and hypothesis testings can be constructed based on this asymptotic normality. For conciseness, we focus on confidence intervals in this work, while a similar procedure can be adopted for hypothesis testings.
Under some mild assumptions on the signal and sample size, we establish theoretical guarantees for the proposed method. These assumptions are generally weak in the sense that the dimension could exceed the sample size and many non-zero small coordinates are allowed. Furthermore, theoretical analysis reveals that the modified estimators for individual coordinates have uniformly bounded variance, and hence simultaneous interval estimation is possible. Numerical simulations in a wide range of settings are supportive of our theoretical results.

\end{abstract}


\section{Introduction}
The problem of recovering a signal from its transformed measurements, referred to as phase retrieval, is fundamental in various applications, including optical imaging, X-ray crystallography, speech recognition, \textit{etc} \cite{shechtman2015phase}. It can be formulated into model (\ref{model}). 
\begin{equation}
   y_j=\mid{\bxj^* \bbeta}\mid^2+\varepsilon_j,\hspace{1cm}j=1,\cdots,n
   \label{model}
\end{equation}
where $\varepsilon_j$ is a random noise with mean zero, $\bxj, \bbeta \in \mathbb{C}^p$ or $\mathbb{R}^p$, and $\bxj^*$ denotes the conjugate transpose of $\bxj$.
Given the noise-contaminated magnitudes $y_j$'s and the design vectors $\bxj$'s, we need to recover the signal $\bbeta$. $\bxj$'s could be Fourier basis, Gaussian vectors, or other sensing vectors, depending on the specific scenario. The problem is difficult because phase information is totally lost in the data-acquisition process. Extensive literature is available on the theory and algorithms for estimating $\bbeta$. 
 
The early-stage algorithms pioneered by Gerchberg and Saxton \cite{gerchberg1972practical} and extended by Fienup \cite{fienup1987reconstruction} start with an arbitrary guess, then refine it by transforming back and forth between the signal domain and Fourier domain until all the constraints are satisfied. Since the violation between the iterate and the $\textit{a priori}$ knowledge is monotonically non-increasing, this type of algorithms get the name error reduction algorithms \cite{fienup1982phase}. Such scheme is equivalent to alternating projections onto nonconvex sets \cite{levi1984image,bauschke2003hybrid}, but its convergence nature remains unknown. Besides, error reduction algorithms rely heavily on the prior information of the signal. Following the spirit of Gerchberg-Saxton algorithm, the alternating minimization is recently proposed \cite{netrapalli2013phase}. It divides the data into a number of independent parts and use a new part in each minimization step. Nevertheless, this strategy is of little practical value.  
 
In most literature, phase retrieval is translated into a nonconvex minimization problem with various objective functions, $\textit{e.g.}$, (\ref{objective1})-(\ref{objective3}).
\begin{equation}
   \underset{\bb \in \mathbb{C}^p/\mathbb{R}^p}{\text{minimize}}
   f(\bb)=\frac{1}{4n}\sum_{j=1}^n (|\bxj^* \bb|^2-y_j)^2
   \label{objective1}
\end{equation}

\begin{equation}
   \underset{\bb \in \mathbb{C}^p/\mathbb{R}^p}{\text{minimize}}
   f(\bb)=\frac{1}{4n}\sum_{j=1}^n (|\bxj^* \bb|-\phi_j)^2,\quad
   where\ \phi_j = |{\bxj^* \bbeta}| + \varepsilon_j 
   \label{objective2}
\end{equation}

\begin{equation}
   \underset{\bb \in \mathbb{C}^p/\mathbb{R}^p}{\text{minimize}}
   f(\bb)=\frac{1}{4n}\sum_{j=1}^n \Big| |\bxj^* \bb|^2-y_j \Big|
   \label{objective3}
\end{equation}

Existing methods for solving (\ref{objective1})-(\ref{objective3}) can be categorized into convex-optimization-type and gradient-descent-type. The former is based on Shor's convex relaxation \cite{ben2001lectures}. It relaxes (\ref{objective1}) to a convex minimization problem (see below) and solves this convex problem via semidefinite programming (SDP) \cite{chai2010array,candes2013phaselift,waldspurger2015phase}. 
\begin{equation*}
\begin{aligned}
&\underset{\bm{B}}{\text{minimize}}
&&\mathrm{trace}(\bm{B}) \\
&\text{subject to}
&&\bm{B} \succeq 0\\
&&&y_j=trace(\bxj\bxj^*\bm{B}),\quad j=1,\cdots,n.
\end{aligned}
\end{equation*}
where $\bm{B}=\bm{b}\bm{b^*}$. 
Under noiseless Gaussian designs, SDP achieves exact recovery with sample size $O(p)$ \cite{candes2013phaselift}. Later, a modified version of SDP is proposed with the trace norm replaced by a reweighted trace norm, which is equivalent to minimizing a log-det function \cite{candes2015phase1, fazel2003log}. Despite its reasonable performance and theoretical guarantees, SDP is computationally expensive because it optimizes over $p^2$ variables.

The second category are various types of gradient descent methods. The "Wirtinger flow" algorithm \cite{candes2015phase} targets the objective function (\ref{objective1}). It obtains the starting point by spectral initialization, and refines the iterates via Wirtinger derivatives. A sample size $O(p\log p)$ is claimed to guarantee reasonable accuracy. The "truncated Wirtinger flow" \cite{chen2015solving} eliminates those abnormal data points generated during the process to obtain a more reliable starting point as well as control the search direction. It exhibits more stable performance than the plain Wirtinger flow algorithm while advances the sample complexity to $O(p)$. The "truncated amplitude flow" algorithm \cite{wang2018solving} is targeting the objective function (\ref{objective2}). It is also a two-stage procedure with orthogonality-promoting initialization followed by regularized gradient descent \cite{wang2018solving}. During its gradient descent stage, the signs of the components $\bxj^*\bb$ are scrutinized to ensure a correct search direction. In a recent paper \cite{duchi2017solving}, the objective function (\ref{objective3}) is transformed into the composition of a convex function and a smooth function, and the smooth function is further approximated by a linear function. The resulting  objective function is convex and amenable to gradient descent. This method has slightly broader applications than other methods because it works on certain complex design vectors besides Gaussian designs. We do not give a complete bibliography here due to the vast amount of literature on this topic. 

In many real-world applications, the signal has few nonzero coordinates, and far less measurements than the dimension of the signal are available. Phase retrieval in such context is referred to as sparse phase retrieval. The community has showed extensive interest on sparse phase retrieval during the past two decades \cite{donoho2006compressed}. Many of the algorithms for sparse phase retrieval are obtained by modifying the existing algorithms for non-sparse case. For example, certain norm regularization is added to the trace function in SDPs to promote sparsity \cite{ohlsson2011compressive, li2013sparse, oymak2015simultaneously}, and a thresholding step is incorporated into each iteration of the gradient-descent-type algorithms \cite{cai2016optimal, wang2017sparse}. These modifications do not work for Fourier phase retrieval due to the ambiguity of translation and conjugate reflection. A novel method is proposed in \cite{jaganathan2012recovery} and demonstrated good performance in sparse Fourier phase retrieval. It first estimates the support via autocorrelation function, and then solves an SDP over the support. Another recent algorithm is based on greedy local search \cite{shechtman2014gespar}. It updates the signal support by interchanging the coordinate on support with the smallest gradient value with the coordinate off support with the largest gradient value. The objective function is also updated accordingly in each iteration. There are some other established methods for (sparse) phase retrieval \cite{jaganathan2013sparse}\cite{iwen2017robust}\cite{jaganathan2015phase}, which we will not elaborate here to avoid unnecessary details. 

Despite such intensive study on the algorithms for solving phase retrieval and sparse phase retrieval, statistical inferences about the signal is rarely touched. All the foregoing methods merely generate a point estimator for $\bbeta$ and establish its convergence rate. No statistical inferences can be drawn on $\bbeta$ or a function of $\bbeta$ based on these point estimators. While in many real life applications, statistical inferences on the sparse signal are very much desired. For example, researchers might seek the $95\%$ confidence interval of a certain coordinate $\beta_k$ in order to adjust the receiver bandwidth. 
One major obstacle that thwarts statistical inferences on sparse phase retrieval is that the estimators generated by these algorithms cannot be written as an explicit function of the data. Thus, its sampling distribution or asymptotic distribution is in general not tractable. 

In this paper, we propose a general method to construct confidence intervals for some simple function of the signals $\theta(\bbeta)$, \textit{e.g.} $\beta_k$. We also show that the resulting confidence interval approximately attains the preassigned coverage probability when the sample size satisfies $n\gg (\log p)^2/s$.
Suppose we have an estimator for $\theta(\bbeta)$ available, which is asymptotically normal with mean $\theta(\bbeta)$. Then confidence intervals can be easily built based on this estimator and its asymptotic normality. Therefore, the key is to construct such an estimator, which we shall obtain as follows. First we pick an initial estimator output from some existing phase-retrieval algorithm; second we modify the initial estimator in a way that the resulting estimator possesses all desired properties (asymptotically normal and centered at the true $\theta(\bbeta)$). The choice for the initial estimator will be discussed in Section 2. 
This method is inspired by the "debiased LASSO" introduced in high-dimensional linear regressions \cite{zhang2014confidence} \cite{van2014asymptotically} \cite{javanmard2014hypothesis} \cite{bellec2019biasing}. The LASSO is a shrinkage/thresholded estimator and hence biased. By adding a bias-correction term to the initial LASSO estimator, the authors obtain an asymptotically unbiased and normally distributed estimator, the debiased LASSO. Similarly in the case of sparse phase retrieval, existing algorithms always generate biased estimators because they are designed to promote sparsity. We will try to eliminate the bias of the chosen initial estimator by adding a different correction term other than that in debiased LASSO. The correction term is core to our method, and will be derived in Section 2. 
To the best of our knowledge, this is the first effort on statistical inference in the realm of sparse phase retrieval. 

We organize the rest of this paper as follows. Section 2 elaborates our methodology and explains the rationale behind it. Section 3 presents the main theoretical guarantees for our method. Section 4 displays its empirical performance. Further discussions and perspectives are left to Section 5. And we leave all the proofs and technicality to Section 6.

\section{Methodology}
Throughout the rest of the paper, we carry out the discussion based on model (\ref{model}) and objective function (\ref{objective1}). Wherever $f(\cdot)$ appears, it means the function in (\ref{objective1}). However, the idea extends naturally to other phase retrieval models and objective functions. For conciseness, we focus on the real case, and consider model (\ref{model}) with $\bbeta \in \mathbb{R}^p, \|\bbeta\|_0=s$, $\bxj$'s i.i.d. $N(\bm{0},I_p)$, and $\varepsilon_1, \varepsilon_2, \cdots, \varepsilon_n \in \mathbb{R}$ i.i.d. Gaussian noise $N(0,\sigma)$. Given the measurements $\y$ and the design matrix $\X$, we aim to construct confidence intervals with approximately preassigned coverage probabilities for a one-dimensional parameter $\theta=\theta(\bbeta)=\beta_k$, one coordinate of the signal.  

This work is motivated by the celebrated debiasing techniques \cite{zhang2011statistical}\cite{zhang2014confidence}\cite{van2014asymptotically}\cite{javanmard2014hypothesis}\cite{bellec2019biasing} proposed for high-dimensional linear regressions and the various phase retrieval algorithms described in Section 1. Provided an initial estimator for $\beta_k$, which is biased and whose sampling distribution is intractable, we correct its bias so that the resulting estimator has approximately normal distribution centered at $\beta_k$ in asymptotic. And confidence intervals can be constructed based on this asymptotic normality. 
Although we restrict our discussion to sparse phase retrieval problem, the method is applicable to any M-estimating problem where (i) the Hessian matrix exits in a big enough neighborhood of the global maximizer and is invertible, (ii) $\theta(\bbeta)$ is differentiable almost everywhere, (iii) a good enough initial estimator is available.

\subsection{Choice of the initial estimator}
Thresholded Wirtinger Flow (TWF) is recently proposed for sparse phase retrieval \cite{cai2016optimal}. It first generates a starting point by spectral initialization and then apply thresholded gradient descent to refine it. We choose the TWF output as our initial estimator because it is claimed to achieve optimal minimax rate of convergence in the sparse phase retrieval setting \cite{cai2016optimal}. More specifically, The authors showed that with high probability the TWF estimator of iteration t, $\tb^{(t)}$, falls within a tiny ball centered at $\bbeta$ (Theorem \ref{thm-A}). 
Written in mathematical formula,
\begin{equation}
     \inf_{\|\bbeta\|_0=s}\mathbb{P}_{(\X,\y|\bbeta)}\bigg\{\min_{i=0,1}\|\tb^{(t)}-(-1)^{i}\bbeta\|_2\leq\frac{1}{6}(1-\frac{\mu}{16})^t\|\bbeta\|_2+C\frac{\sigma}{\|\bbeta\|_2}\sqrt{\frac{s\log{p}}{n}}\bigg\} > 1-\frac{46}{n}-10e^{-s}-\frac{t}{np^2}
     \label{TWF1}
\end{equation}
for some absolute constant $C > 0$, provided the sample size $n\geq K(1+\frac{\sigma}{\|\bbeta\|_2^2})^2 s^2\log(np)$ for some absolute constant $K > 0$ and the tuning parameters in the TWF algorithm are properly chosen. Here $\mu$ is the step size of gradient descent, which can be regarded as absolute constant once it is decided. If $t\asymp \log\bigg(\frac{\|\bbeta\|_2^2\sqrt{n}}{\sigma\sqrt{s\log p}}\bigg)$, one can obtain from the preceding result that with high probability
\begin{equation}
    \min_{i=0,1}\|\tb^{(t)}-(-1)^{i}\bbeta\|_2\precsim \frac{\sigma}{\|\bbeta\|^2_2}\sqrt{\frac{s\log p}{n}}.
    \label{TWF2}
\end{equation}
This error rate is crucial for our bias-correction scheme to work, as will be revealed later in the proof part. 

\subsection{Bias-correction strategy}
In this subsection, we will explain the bias-correction procedure in details. Suppose we are interested in a one-dimensional parameter $\theta=\theta(\bbeta)$, which is a continuously differentiable function of $\bbeta$. In our particular case, $\theta=\bm{e}_j^T\bbeta=\beta_j,\ j=1,2,\cdots,p$. The TWF solution, denoted by $\tb$, is biased due to the shrinkage nature of TWF. And so is $\theta(\bbeta)$.
To correct the bias of $\theta(\tb)$, we adopt the idea of Low-Dimensional Projection Estimator (LDPE) proposed in \cite{zhang2011statistical}. The authors consider a more general semi-low-dimensional (LD) approach where a high-dimensional (HD) model is decomposed as 
\begin{equation*}
    HD model = LD component + HD component
\end{equation*}
In our problem, such decomposition amounts to 
\begin{equation}
    \bbeta-\tb=\bu(\tb) \bigg(\theta(\bbeta)-\theta(\tb)\bigg) + Q(\tb)(\bbeta-\tb)
    \label{decomposition}
\end{equation}
Here $\bu(\tb)$ is the least favorable one-dimensional sub-model at $\tb$, normalized such that $\nabla\theta(\tb)^T\bu(\tb)=1$. It is called least favorable because it gives the minimum Fisher information for estimating $\theta$. The formula for $\bu(\tb)$ is given by
\begin{equation*}
    \bu(\tb) = \frac{\I(\tb)^{-1} \nabla\theta(\tb)}{\nabla\theta(\tb)^T\I(\tb)^{-1}\nabla\theta(\tb)}
\end{equation*}
where $\I(\tb)$ is the Fisher Information at $\tb$ 
\begin{equation*}
    \I(\tb) := - \mathbb{E}_{\tb}\Bigg[\frac{\partial^2 l(\X,\y|\bb)}{\partial \bb \partial \bb^T}\Bigg]_{\bb=\tb}
\end{equation*}
While $Q(\tb)$ projects $\bbeta-\tb$ into a space of nuisance parameters. To simplify the notation, let $\tu=\bu(\tb)$. The LDPE searches, in the least favorable direction $\tu$, a parameter value that maximizes the likelihood of the occurring data sample. 
\begin{equation}
    \hat{\theta} = \theta(\tb) + \argmax_{d}\sum_{j=1}^n l(\bxj,y_j|\tb + \tu d) = \theta(\tb) + \argmin_{d}f(\tb + \tu d)
    \label{MLE}
\end{equation}
The second equality above is because under model (\ref{model}) and Gaussian noises, 
\begin{equation*}
    \sum_{j=1}^n l(\bxj,y_j|\bb) \propto \frac{1}{4n}\sum_{j=1}^n -(y_j-|\bxj^T\bb|^2)^2 = -f(\bb)
\end{equation*}

The score equation for the minimization problem in (\ref{MLE}) is 
\begin{equation*}
    \tu^T \nabla f(\tb+\tu \hat{d}) = 0,
\end{equation*}
Where $\hat{d}=\argmin_{d}f(\tb + \tu d)$. 
By Taylor expansion, 
\begin{align*}
    \nabla f(\tb+\tu \hat{d}) - \nabla f(\tb) &\approx \hat{d}\Bigg[\frac{\partial^2 f(\bb)}{\partial \bb \partial \bb^T}\Bigg]_{\bb=\tb}\tu \\
    &\approx \hat{d}\I(\tb)\tu \\
    &= \hat{d}\frac{\nabla\theta(\tb)}{\nabla\theta(\tb)^T\I(\tb)^{-1}\nabla\theta(\tb)}
\end{align*}
Thus we have 
\begin{align*}
    0=\tu^T \nabla f(\tb+\tu \hat{d}) &\approx \tu^T \nabla f(\tb) + \hat{d} \frac{\tu^T\nabla\theta(\tb)}{\nabla\theta(\tb)^T\I(\tb)^{-1}\nabla\theta(\tb)} \\
    &= \tu^T \nabla f(\tb) + \frac{\hat{d}}{\nabla\theta(\tb)^T\I(\tb)^{-1}\nabla\theta(\tb)}
\end{align*}
which solves
\begin{equation}
    \hat{d}\approx - \nabla\theta(\tb)^T \I(\tb)^{-1}\nabla f(\tb)
    \label{hatd}
\end{equation}

Plug (\ref{hatd}) into (\ref{MLE}), our corrected estimator is given by
\begin{equation}
   \hat{\theta} \approx \theta(\tb) - \nabla\theta(\tb)^T \I(\tb)^{-1}\nabla f(\tb)
   \label{debiased1}
\end{equation}

Later we will implement the bias-corrected estimator (\ref{debiased1}) to the case where $\theta(\bbeta)=\ek^T\bbeta=\beta_k$, and demonstrate some nice properties of this estimator. For consistency, hereinafter we will denote the corrected estimator as $\hat{\bbeta}$ and the initial estimator as $\tb$. 

Before elaborating the analysis, let us acknowledge one fact. The exact sign of an individual $\beta_k$ cannot be recovered because the measurements only provide magnitude information ($|\bxj^T\bbeta|^2$ contaminated with noise). If $\tb$ is a solution to the minimization problem (\ref{objective1}), $-\tb$ is also a solution. The signal $\bbeta$ can be recovered only up to a global sign. Given an initial estimator $\tb$, we define $\bs$ to be whichever in $\{\bbeta, -\bbeta\}$ is closer to $\tb$.
\begin{equation*}
    \bs := \left\{
    \begin{array}{ccc}
         \bbeta & \text{if} \ \|\bbeta-\tb\|^2_2 \leq \|-\bbeta-\tb\|^2_2 \\
         -\bbeta & \text{otherwise} \\
    \end{array}
    \right.
\end{equation*}
We are only interested in gaining information about $\bs$. This is the best we can do under model (\ref{model}) in a sense that the global sign can never be retrieved.

\bigbreak
To prepare the proofs in Section 6, we derive some formulae and simply a few notations here. 
The Fisher information matrices at $\bs$ and $\tb$ are 
\begin{align}
    \I(\bs) &= \mathbb{E}\big[2(\bxj^T\bs)^2\bxj\bxj^T-\varepsilon_j\bxj\bxj^T\big] = \lnorm^2 I+2\bs {\bs}^T \nonumber\\
    \I(\tb) &= \mathbb{E}\big[2(\bxj^T\tb)^2\bxj\bxj^T-\varepsilon_j\bxj\bxj^T\big] = \tlnorm^2 I+2\tb \tb^T
    \label{fisherinf}
\end{align}
In our specific context, $\theta(\bs)=\ek^T\bs$ and $\nabla\theta(\bs) = \ek$. If we define
\begin{align}
    \wk := w(\bs, \theta=\beta_k) = -\frac{1}{2}\I(\bbeta)^{-1}\nabla \theta(\bbeta) &= -\frac{1}{2}\big(\|\bs\|^2_2 I+2\bs {\bs}^T\big)^{-1}\ek \nonumber\\
    &= -\frac{1}{2\|\bs\|_2^2} \bigg( I-\frac{2\bs {\bs}^T}{3\|\bs\|_2^2} \bigg)\ek,
    \label{wtrue}
\end{align}
and
\begin{align}
    \twk := w(\tb, \theta=\beta_k) = -\frac{1}{2}\I(\tb)^{-1}\nabla \theta(\tb) &= -\frac{1}{2}\big(\|\tb\|^2_2 I+2\tb \tb^T\big)^{-1}\ek \nonumber\\
    &= -\frac{1}{2\|\tb\|_2^2} \bigg( I-\frac{2\tb {\tb}^T}{3\|\tb\|_2^2} \bigg)\ek,
    \label{west}
\end{align}
the estimator (\ref{debiased1}) can adopt a simpler formula
\begin{equation}
    \hat{\beta}_k = \Tilde{\beta}_k + \twk^T \nabla f(\tb)
    \label{debiased2}
\end{equation}

$\bm{Remark\ 2.2.}$ Equation (\ref{fisherinf}) is derived under the assumption that $\tb$ is independent from the design vectors $\{\bxj\}_{j=1,\cdots,n}$. This assumption is realistic in certain specially-designed procedure. For example, we can randomly split the i.i.d. data ${(\bxj, y_j)}_{j=1,\cdots,n}$ into two parts. We use the first part to obtain $\tb$ and the second part to construct the bias-correction term. The detailed procedure will be addressed in Section 2.3.

\subsection{Data split and swap}
In the analysis so far, we implicitly assume that $\tb$ is independent from the design vectors $\bxj$'s and the noises $\varepsilon_j$'s, which especially ease the derivation of (\ref{fisherinf}). However, this assumption fails if $\tb$ is obtained from the same set of design vectors $\bxj$'s and measurements $y_j$'s by TWF iterations. To validate the independence between $\tb$ and ($\{\bxj\}_j$, $\{\varepsilon_j\}_j$), we design the three-step procedure. First, splitting the whole data set randomly into two parts, using the first part of data $(\X_1, \y_1)$ to obtain $\tb$ via TWF and using the second part of data $(\X_2, \y_2)$ to debias $\tb$; second, swapping the two parts of data, using the second part to generate $\tb$ and the first part to debias $\tb$; finally, combining the two debiased estimators. Theoretically, the required sample size to obtain a good initial estimator $\tb$ is different from the required sample size for efficient bias-correction. But we set the two parts to be of equal size n since the two parts will be swapped for both purposes.

\section{Theoretical Result}
In this section, we will establish the asymptotic normality of the bias-corrected estimator (\ref{debiased2}). And the confidence interval based on this asymptotic distribution is shown to achieve roughly the preassigned coverage probability. Furthermore, theoretical analysis reveals that the confidence interval has sharp width and attains approximately the coverage probability simultaneously for all $\beta_k$. 
Note we assume $\sigma$ is known because in many applications people have $\textit{a prior}$ knowledge about the noise level. Even if the noise level is unknown, there are efficient ways to estimate it, for example, the method proposed in \cite{cai2016optimal}.

Theorem \ref{thm-1} provides theoretical guarantees for the data-splitting scheme described below. The $\textit{i.i.d.}$ data generated from model (\ref{model}) are randomly split into two halves $(\X_1, \y_1)$, $(\X_2, \y_2)$. We use $(\X_1, \y_1)$ to obtain the initial estimator and $(\X_2, \y_2)$ to correct the bias. To fully extract information contained in the data, we further apply the data-swapping scheme, whose theoretical property is given in Theorem 3.2. More precisely, it goes through two rounds of construction, generating a bias-corrected estimator in each round. In the first round, we obtain an initial estimator from $(\X_1, \y_1)$ and correct bias using $(\X_2, \y_2)$. In the second round, the initial estimator is generated from $(\X_2, \y_2)$ while $(\X_1, \y_1)$ is used to correct the bias. Finally we combine the two estimators in a way that the resulting estimator has the smallest asymptotic variance. Each half of the data has size n.

Before stating the theorems, we introduce some global assumptions. Suppose the design matrix and the true signal satisfy (\ref{assumption1})-(\ref{assumption3}).
\begin{equation}
    n \geq K(1+\frac{\sigma}{\lnorm^2})^2 s^2\log(np), \quad K\ is\ an\ absolute\ constant.
    \label{assumption1}
\end{equation}
   
\begin{equation}
    \frac{\log p}{\sqrt{ns}} = o(1).
    \label{assumption2}
\end{equation}
    
\begin{equation}
    \lnorm = O(\sqrt{s}).
    \label{assumption3}
\end{equation}
Assumption (\ref{assumption1}) is to guarantee the quality of the initial estimator, the TWF estimator, as stated in \cite{cai2016optimal}. For the sack of bias-correction, assumptions (\ref{assumption2}) and (\ref{assumption3}) would be sufficient. If $s\ll \log p$, assumptions (\ref{assumption2}) and (\ref{assumption3}) would imply assumption (\ref{assumption1}). However, we do not impose here any restrictions on $s$. Our method could work for cases where the signal sparsity is not strong, \textit{i.e.}, there could be many non-zero small coordinates.   

Throughout the rest of the paper, let 
\begin{equation}
    \epsilon_n'= C_1\frac{(\log p)^{\frac{3}{2}}}{sn} + C_2\frac{(\log p)^2}{(sn)^{\frac{3}{2}}} + C_3\frac{\log p}{(sn)^\frac{1}{2}} + C_4\frac{n^{\frac{1}{2}}}{p^2}
    \label{eps-single}
\end{equation}
for some absolute constants $C_1$, $C_2$, $C_3$, $C_4$, and let
\begin{equation}
    \epsilon_n''=\frac{50}{n}+10e^{-s}+\frac{M\log n}{np^2}
    \label{eps-double}
\end{equation}
for some  absolute constant M.
It is easy to see that $\epsilon_n', \epsilon_n''\rightarrow\ 0$ under assumption (\ref{assumption2}), (\ref{assumption3}).
\bigbreak 

\begin{theorem}\label{thm-1}
If we randomly split the data into two halves, using the first half to obtain the TWF estimator with enough iterations such that $\|\tb-\bs\|_2\leq \frac{C_0\sigma}{\lnorm}\sqrt{\frac{s\log p}{n}}$, and the second half to construct the bias-correction term, then
\begin{equation}
    \mathbb{P}\Bigg\{\left\lvert\sqrt{n}(\hat{\beta}_k-\beta_k^*)-Z_k \right\rvert \leq \epsilon_n'\Bigg\} \geq 1-\epsilon_n''
    \label{thm1}
\end{equation}
with
\begin{equation}
    Z_k=-\frac{1}{\sqrt{n}}\sum_{j=1}^n \varepsilon_j(\bxj^T\tb)(\bxj^T\twk),
    \label{Zk}
\end{equation}
where $\hat{\beta}_k$, $\twk$, $\epsilon_n'$, $\epsilon_n''$ are given by (\ref{debiased2}), (\ref{west}), (\ref{eps-single}), (\ref{eps-double}), respectively.
\end{theorem}

\bigbreak
$\bm{Remark\ 3.1}$ It is clear that $\tb$ is independent from $\bxj$'s in our data-splitting regime.
$Z_k$ has limiting distribution $N(0,\sigma^2\tau_k^2)$, where
\begin{align*}
    \tau_k^2 &=\tlnorm^2\|\twk\|_2^2+2(\tb^T\twk)^2 =\tlnorm^2\|\twk\|_2^2 + 2\bigg(-\frac{\Tilde{\beta}_k}{2\tlnorm^2}+\frac{\Tilde{\beta}_k}{3\tlnorm^2}\bigg)^2\\
    &=\tlnorm^2\|\twk\|_2^2 +\frac{\Tilde{\beta}_k^2}{18\tlnorm^4}\nonumber\\
    &\leq \bigg(\lnorm+\|\tb-\bs\|_2\bigg)^2\bigg[\frac{5}{6\lnorm^2}+ O(\frac{\|\tb-\bs\|_2}{\lnorm^3}) \bigg]^2 + \frac{1}{18\tlnorm^2} \asymp \frac{3}{4s} + O(\sqrt{\frac{\log p}{n s^3}})
\end{align*}
by (\ref{assumption3}), (\ref{west}), (\ref{TWF2}) and (3.19).
Such sharp width is uniformly achievable for all coordinates, which allows the possibility of constructing simultaneous confidence intervals. 

\bigbreak
To fully utilize the data, we further carry out the data-swap procedure. It is composed of two rounds of data-splitting procedures: dividing the data randomly into two halves $(\X_1, \y_1)$ and $(\X_2, \y_2)$; in the first round, we obtain the initial estimator $\tb_1$ from $(\X_1, \y_1)$, and correct the bias using $(\X_2, \y_2)$, resulting in $\hat{\beta}_{1k}=\Tilde{\beta}_{1k}+\ftwk^T\nabla f_2(\tb_1)$; in the second round, we obtain the initial estimator $\tb_2$ from $(\X_2, \y_2)$ and correct the bias using $(\X_1, \y_1)$, resulting in $\hat{\beta}_{2k}=\Tilde{\beta}_{2k}+\stwk^T\nabla f_1(\tb_2)$. Here
\begin{align*}
    \ftwk:= w(\tb_1,\theta=\beta_k)=-\frac{1}{2}\bigg(\|\tb_1\|_2^2I+ 2\tb_1\tb_1^T\bigg)^{-1}\ek\\
    \stwk:= w(\tb_2,\theta=\beta_k)=-\frac{1}{2}\bigg(\|\tb_2\|_2^2I+ 2\tb_2\tb_2^T\bigg)^{-1}\ek\\
    \nabla f_2(\tb_1)=\frac{1}{n}\sum_{j=1}^n \big[(\sbxj^T\tb_1)^2-y_{2j}\big](\sbxj^T\tb_1)\sbxj\\
    \nabla f_1(\tb_2)=\frac{1}{n}\sum_{j=1}^n \big[(\fbxj^T\tb_2)^2-y_{1j}\big](\fbxj^T\tb_2)\fbxj
\end{align*}
Finally, we linearly combine $\hat{\beta}_{1k}$ and $\hat{\beta}_{2k}$, resulting in a better estimator $\hat{\beta}_k^{swap}$ that contains more information of the data than $\hat{\beta}_{1k}$ or $\hat{\beta}_{2k}$ alone.
\bigbreak

\begin{theorem}\label{thm-2}
$\bm{(i)}$\quad Under the global assumptions (\ref{assumption1})-(\ref{assumption3}), the estimator 
\begin{equation}
    \hat{\beta}_k^{swap}=\frac{\tau_{2k}^2}{\tau_{1k}^2+\tau_{2k}^2}\cdot\Bigg(\Tilde{\beta}_{1k}+\ftwk^T\nabla f_2(\tb_1)\Bigg) + \frac{\tau_{1k}^2}{\tau_{1k}^2+\tau_{2k}^2}\cdot\Bigg(\Tilde{\beta}_{2k}+\stwk^T\nabla f_1(\tb_2)\Bigg)
    \label{beta_swap}
\end{equation}
achieves the smallest asymptotic variance among all asymptotically unbiased estimators that are convex combinations of $\hat{\beta}_{1k}$ and $\hat{\beta}_{2k}$.
Furthermore, 
\begin{align}
     \mathbb{P}\Bigg\{\left\lvert\sqrt{n}(\hat{\beta}^{swap}_k-\beta_k^*)-\bigg(\frac{\tau_{2k}^2}{\tau_{1k}^2+\tau_{2k}^2} Z_{1k} + \frac{\tau_{1k}^2}{\tau_{1k}^2+\tau_{2k}^2} Z_{2k}\bigg) \right\rvert \leq \epsilon_n'\Bigg\} \geq 1-2\epsilon_n''
     \label{thm2_1}
\end{align}
where
\begin{align*}
    Z_{1k} =\frac{1}{\sqrt{n}}\sum_{j=1}^n \varepsilon_{2j}(\sbxj^T\tb_1)(\sbxj^T\ftwk)\\
    Z_{2k} =\frac{1}{\sqrt{n}}\sum_{j=1}^n \varepsilon_{1j}(\fbxj^T\tb_2)(\fbxj^T\stwk)\\
    \tau_{1k}^2 = \|\tb_1\|_2^2\|\ftwk\|_2^2+2(\tb_1^T\ftwk)^2\\
    \tau_{2k}^2 = \|\tb_2\|_2^2\|\stwk\|_2^2+2(\tb_2^T\stwk)^2
\end{align*}
$\bm{(ii)}$\quad 
\begin{align}
     \mathbb{P}\Bigg\{\left\lvert \hat{\beta}^{swap}_k-\beta_k^*\right\rvert \leq \frac{\sigma r}{\sqrt{n}}\sqrt{\frac{\tau_{1k}^2\tau_{2k}^2}{\tau_{1k}^2+\tau_{2k}^2} + O(\frac{1}{p^2})}+ \frac{\epsilon_n'}{\sqrt{n}}\Bigg\} \geq 2\Phi(r)-1-2\epsilon_n''-\frac{2}{n}
     \label{thm2_2}
\end{align}
$\bm{(iii)}$\quad 
\begin{equation}
    \liminf\limits_{n\rightarrow\infty}\mathbb{P}\Bigg\{\max_{k\in [p]}\left\lvert \sqrt{n}(\hat{\beta}^{swap}_k-\beta_k^*)\right\rvert \leq \sqrt{\frac{3}{8s}}\sigma \Phi^{-1}(1-\frac{\alpha}{2})\Bigg\} \geq 1-\alpha
    \label{thm2_3a}
\end{equation}
\begin{equation}
    \liminf\limits_{n\rightarrow\infty}\mathbb{P}\Bigg\{  \left\lvert \bm{h}^T(\hat{\bbeta}^{swap}-\bs)\right\rvert \leq \sqrt{\frac{\sigma^2}{n}\chi^2_{p,\alpha} {\bm{h}^T \bm{V} \bm{h}}}\Bigg\} \geq 1-\alpha\ for\ \forall \bm{h}\neq 0, \bm{h}\in \mathbb{R}^p.
    \label{thm2_3b}
\end{equation}
where 
\begin{equation*}
    V_{kl}=a_k a_l \bigg[\|\tb_1\|_2^2(\ftwk^T\ftwl)+ 2(\tb_1^T\ftwk)(\tb_1^T\ftwl)\bigg] + (1-a_k)(1-a_l) \bigg[\|\tb_2\|_2^2(\stwk^T\stwl)+ 2(\tb_2^T\stwk)(\tb_2^T\stwl)\bigg]
\end{equation*}
 $a_k=\frac{\tau_{2k}^2}{\tau_{1k}^2+\tau_{2k}^2}$, $a_l=\frac{\tau_{2l}^2}{\tau_{1l}^2+\tau_{2l}^2}$.
\end{theorem}

\bigbreak
$\bm{Remark\ 3.2}$\\
Note, asymptotically, $\tau_{1k},\ \tau_{2k} \asymp \frac{3}{4s}$ for all $k=1,\cdots,p$ by Remark 3.1. Hence $\frac{\tau_{1k}^2\tau_{2k}^2}{\tau_{1k}^2+\tau_{2k}^2} \leq \frac{1}{2}\max\{\tau_{1k}^2, \tau_{2k}^2\}\leq \frac{3}{8s}$. The asymptotic variance of $\hat{\beta}_k^{swap}$ is shrunken by a factor of 2 compared to that of $\hat{\beta}_{1k}$ or $\hat{\beta}_{2k}$. The uniformly bounded variance for all $\sqrt{n}(\hat{\beta}^{swap}_k-\beta_k^*)$ in (\ref{thm2_3a}) allows Bonferroni adjustment to control familywise error rate in simultaneous interval estimation. While (\ref{thm2_3b}) provides the theoretical guarantee for Scheffe's simultaneous confidence interval. 

\bigbreak
\bigbreak

\section{Numerical Simulation}
In this section, we implement our method on a variety of settings to assess its empirical performance. Moreover, by comparing the results in different combinations of sparsity (s), sample size (2n), and noise-to-signal ratio (NSR), we get a general idea how the performance depends on those factors. Throughout our simulation, the signal dimension p=1000 and all the tuning parameters in the TWF algorithm are fixed. In each choice of (n,s,NSR), we generate the signal $\bbeta$ by randomly picking the support and assigning nonzero coordinates i.i.d. $N(0,1)$. Given this $\bbeta$, the following procedure is repeated independently for 100 times: first, generate 2n random vectors $\bxj$'s i.i.d. $N(0,I)$, 2n noises $\varepsilon_j$'s i.i.d. $N(0, \sigma^2)$, and the measurements $y_j$'s by (1.1); second, obtain the TWF estimator using the whole dataset $(\X, \y)$ and record the errors $\Tilde{\beta}_k - \beta_k^*$ of four large coordinates ($|\beta_k^*|\approx3$), four median coordinates ($|\beta_k^*| \approx1$), and four small coordinates ($|\beta_k^*| \approx 0.1$), respectively; third, implement the data-swap scheme, obtain the debiased TWF $\hat{\bbeta}^{swap}$, and record the errors  $\hat{\beta}_k^{swap}-\beta_k^*$ of four large/median/small coordinates, respectively. Every summary statistic in table 1 comes from a pool of 400 errors and each histogram in Figure 1/2/3 presents the distribution of a pool of 400 errors. 

The performance of our method is assessed in several aspects, including biasness (Table 1), variance (Table 1), asymptotic normality (Figure 1/2/3), and coverage probability (Table 2).

\begin{table}[!htbp]
\begin{tabular}{lll|l|ll|ll|ll}
\hline
     &     &     &      & \multicolumn{2}{l|}{large coordinates} & \multicolumn{2}{l|}{median coordinates} & \multicolumn{2}{l|}{small coordinates} \\ \cline{5-10} 
n    & s   & NSR &      & TWF                & de-TWF            & TWF                & de-TWF             & TWF                & de-TWF            \\ \hline
3000 & 50  & 0.3 & bias & 0.0108             & 0.0016            & 0.0421             & 0.0017             & -0.0607            & 0.0020            \\
     &     &     & sd   & 0.0174             & 0.0203            & 0.0186             & 0.0216             & 0.0188             & 0.0226            \\
     &     &     & mae  & 0.0144             & 0.0135            & 0.0415             & 0.0155             & 0.0604             & 0.0151            \\ \hline
3000 & 100 & 0.3 & bias & 0.0367             & -0.0061           & 0.0573             & -0.0054            & -0.0872            & 0.0054            \\
     &     &     & sd   & 0.0295             & 0.0838            & 0.0291             & 0.0540             & 0.0173             & 0.0513            \\
     &     &     & mae  & 0.0375             & 0.0316            & 0.0567             & 0.0311             & 0.0987             & 0.0353            \\ \hline
3000 & 150 & 0.3 & bias & -0.0370            & -0.0359           & -0.0928            & 0.0475             & 0.0975             & 0.0025            \\
     &     &     & sd   & 0.0483             & 0.1495            & 0.0498             & 0.1287             & 0.0084             & 0.1346            \\
     &     &     & mae  & 0.0442             & 0.0838            & 0.0921             & 0.0992             & 0.1000             & 0.0939            \\ \hline
3000 & 200 & 0.3 & bias & 0.1007             & 0.2577            & 0.1197             & -0.0474            & -0.0989            & 0.0153            \\
     &     &     & sd   & 0.0679             & 0.3521            & 0.0681             & 0.2335             & 0.0068             & 0.2175            \\
     &     &     & mae  & 0.0976             & 0.2857            & 0.1209             & 0.1630             & 0.1000             & 0.1523            \\ \hline
2000 & 100 & 0.3 & bias & 0.0481             & -0.0271           & 0.0844             & -0.0128            & -0.0972            & -0.0143           \\
     &     &     & sd   & 0.0457             & 0.4064            & 0.0475             & 0.2029             & 0.0091             & 0.1421            \\
     &     &     & mae  & 0.0503             & 0.1264            & 0.0840             & 0.1179             & 0.1000             & 0.0986            \\ \hline
4000 & 100 & 0.3 & bias & -0.0216            & -0.0021           & -0.0499            & 0.0027             & 0.0780             & -0.0044           \\
     &     &     & sd   & 0.0247             & 0.0311            & 0.0252             & 0.0331             & 0.0205             & 0.0350            \\
     &     &     & mae  & 0.0255             & 0.0223            & 0.0498             & 0.0241             & 0.0798             & 0.0248            \\ \hline
5000 & 100 & 0.3 & bias & 0.0288             & 0.0018            & 0.0362             & -0.0017            & -0.0625            & 0.0037            \\
     &     &     & sd   & 0.0193             & 0.0227            & 0.0192             & 0.0233             & 0.0190             & 0.0236            \\
     &     &     & mae  & 0.0282             & 0.0156            & 0.0362             & 0.0154             & 0.0624             & 0.0159            \\ \hline
3000 & 100 & 0.2 & bias & -0.0229            & 0.0078            & -0.0410            & -0.0014            & 0.0662             & -0.0050           \\
     &     &     & sd   & 0.0196             & 0.0848            & 0.0210             & 0.0521             & 0.0209             & 0.0380            \\
     &     &     & mae  & 0.0242             & 0.0221            & 0.0408             & 0.0251             & 0.0678             & 0.0243            \\ \hline
3000 & 100 & 0.4 & bias & -0.0346            & -0.0083           & -0.0827            & 0.0140             & 0.0953             & -0.0081           \\
     &     &     & sd   & 0.0425             & 0.0594            & 0.0418             & 0.0585             & 0.0137             & 0.0634            \\
     &     &     & mae  & 0.0380             & 0.0420            & 0.0841             & 0.0425             & 0.1000             & 0.0412            \\ \hline
3000 & 100 & 0.5 & bias & -0.0462            & -0.0076           & -0.0980            & 0.0141             & 0.0972             & -0.0082           \\
     &     &     & sd   & 0.0480             & 0.0630            & 0.0507             & 0.0685             & 0.0097             & 0.0707            \\
     &     &     & mae  & 0.0494             & 0.0445            & 0.0965             & 0.0438             & 0.1000             & 0.0487            \\ \hline
\end{tabular}
\caption{Summary statistics of TWF errors and debiased TWF errors for large $\beta_k$, median $\beta_k$, and small $\beta_k$ under various simulation settings. "mae" stands for "median absolute error".}
\end{table}

Judging from Table 1, debiased TWF achieves close-to-zero average bias at the cost of slightly larger variance than TWF in all settings except when sparsity s is too large (s=200) or sample size 2n is too small (n=2000). The reason behind such phenomenon could be as follows. The bias-correction term with non-zero mean is supposed to neutralize the bias of TWF. Though it brings extra variance, the amount of this extra variance is negligible as long as n is large enough. Nevertheless, when n is small or s is large, the bias-correction term does not concentrate tightly around its mean. Instead of neutralizing the bias of TWF, it adds extra bias and much larger variance. Table 1 also exhibits a parallel trend between TWF and debiased TWF that their bias and variance get larger as s/NSR increases or n decreases. This is because the performance of debiased TWF depends on the quality of TWF while TWF gets worse as s/NSR increases or n decreases, which has been demonstrated in the original paper \cite{cai2016optimal}. Although there are two abnormal cases (s=200 and n=2000) in Table 1 where the debiased TWF fails, our main theory (Theorem 3.2) is not violated because the $(s,n,p)$ in these cases are largely deviated from assumption (\ref{assumption1}) and (\ref{assumption2}).

Figure 1, 2, and 3 demonstrate the unbiasedness and approximate normality of the debiased TWF in all settings. Figure 1 explores the relationship between the sparsity s and the quality of debiased TWF by fixing p=1000, n=3000, NSR=0.3 while varying s=50, 100, 150, 200 from left to right. The normality and unbiasedness hold better in small s. Besides, the distributions spread wider as s increases in every row. Such phenomenon is not surprising because the asymptotic unbiasedness and normality of $\hat{\beta}_k^{swap}$ rely on that $\frac{\sqrt{n}\dnorm^2}{\lnorm}$ is asymptotically negligible. 
When s increases to an extend that the assumption is violated, the quality of $\tb$ drops and $\dnorm=\|\tb-\bs\|_2$ cannot be controlled. Similarly, the errors of debiased TWF distribute more Gaussian and spread narrower as n increases in Figure2 or as NSR decreases in Figure 3.

\newgeometry{left=1in,right=1in,top=0.6in,bottom=0.8in}
\begin{figure}[!htbp]
\centering
\begin{subfigure}[b]{1\textwidth}
   \centering
   \includegraphics[width=0.7\linewidth]{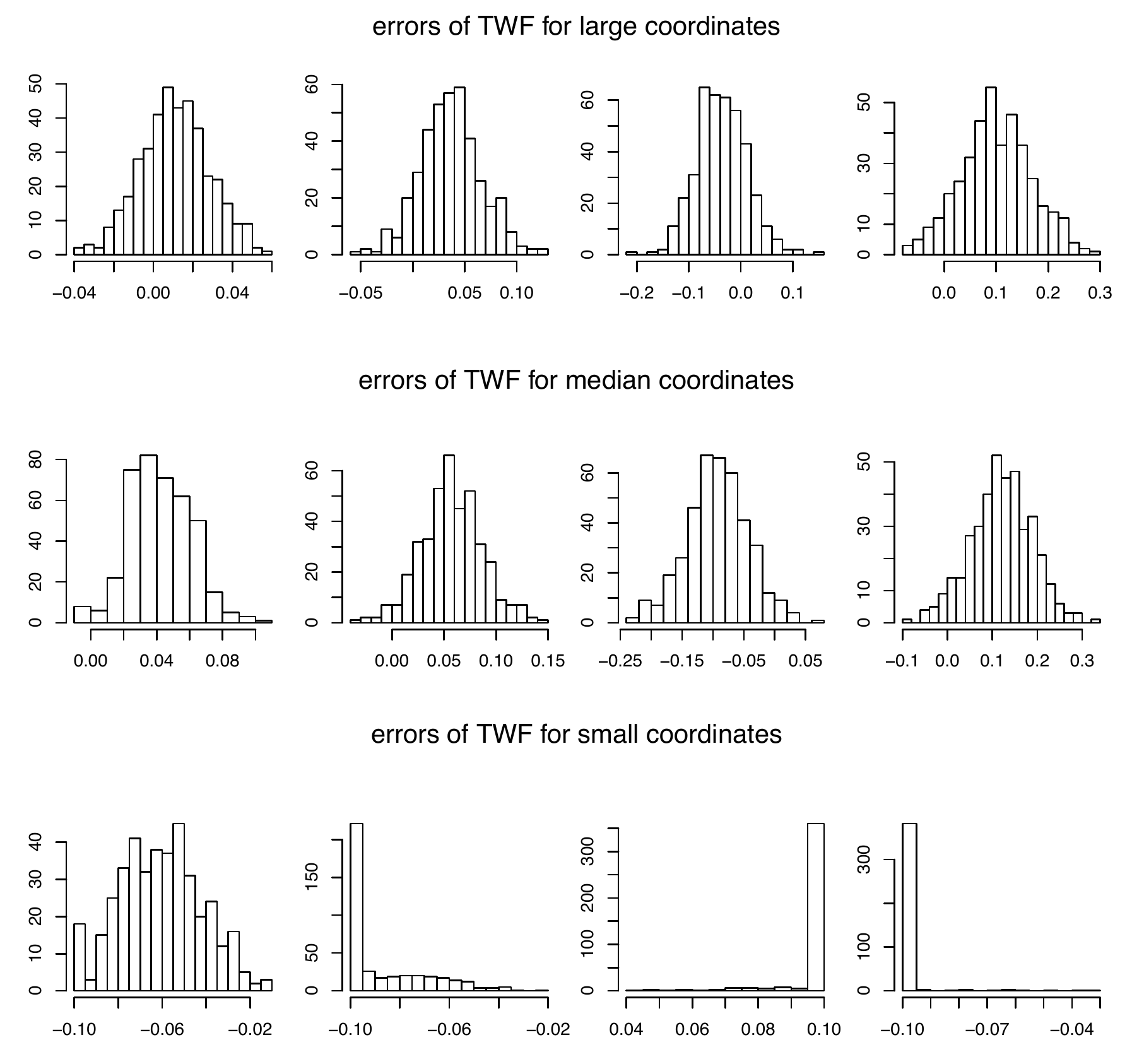}
   \caption{Histograms of TWF errors for large $\beta_k$ (1st row), median $\beta_k$ (2nd row), small $\beta_k$ (3rd row).}
   \label{}
\end{subfigure}
\begin{subfigure}[b]{1\textwidth}
   \centering
   \includegraphics[width=0.7\linewidth]{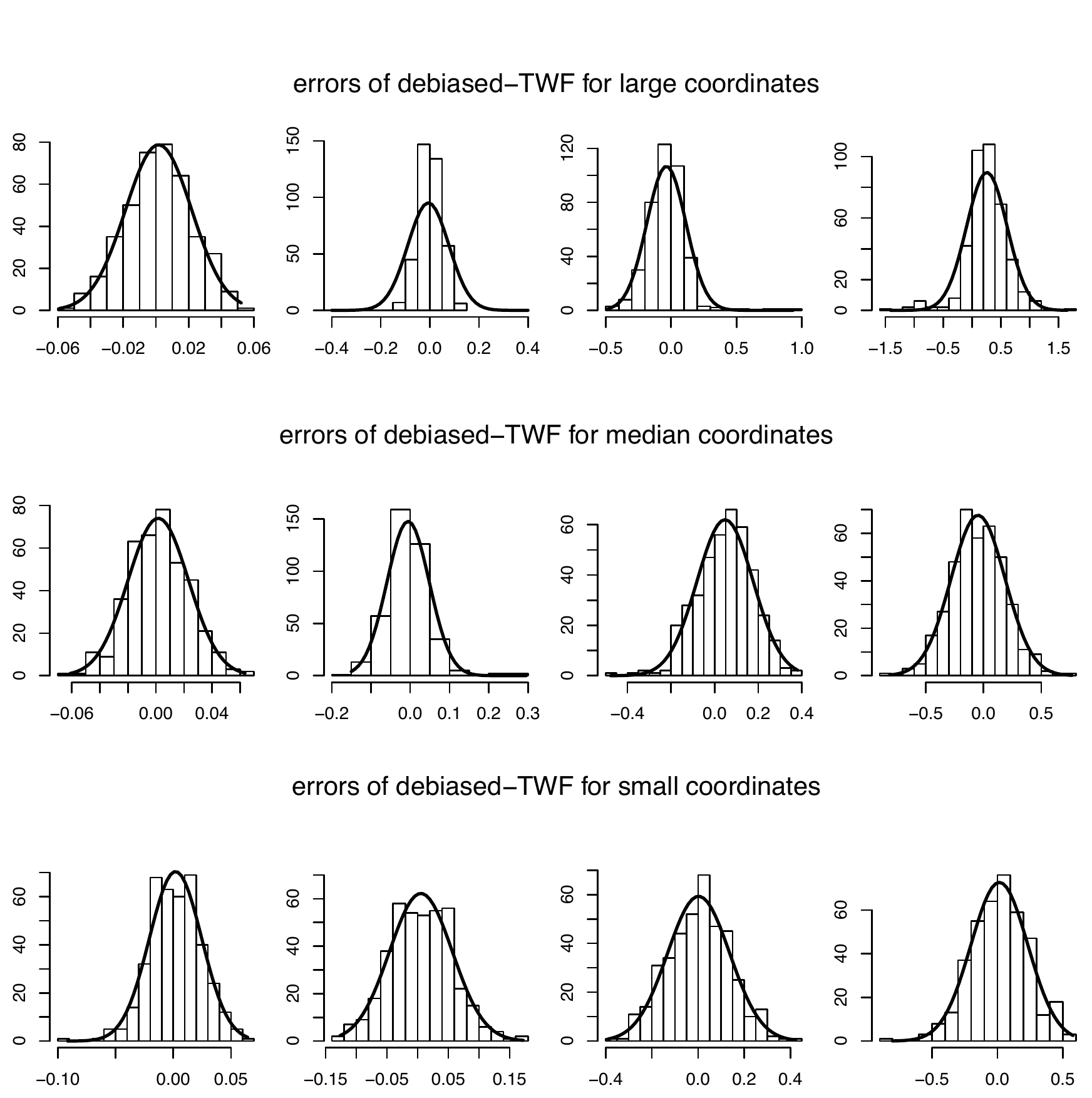}
   \caption{Histograms of debiased TWF errors for large $\beta_k$ (4th row), median $\beta_k$ (5th row), small $\beta_k$ (6th row). }
\end{subfigure}

\caption{ From left to right, plots correspond to simulation settings (n=3000, s=50, NSR=0.3), (n=3000, s=100, NSR=0.3), (n=3000, s=150, NSR=0.3), (n=3000, s=200, NSR=0.3) with p=1000 fixed.}
\end{figure}

\begin{figure}[!htbp]
\centering
\begin{subfigure}[b]{1\textwidth}
   \centering
   \includegraphics[width=0.7\linewidth]{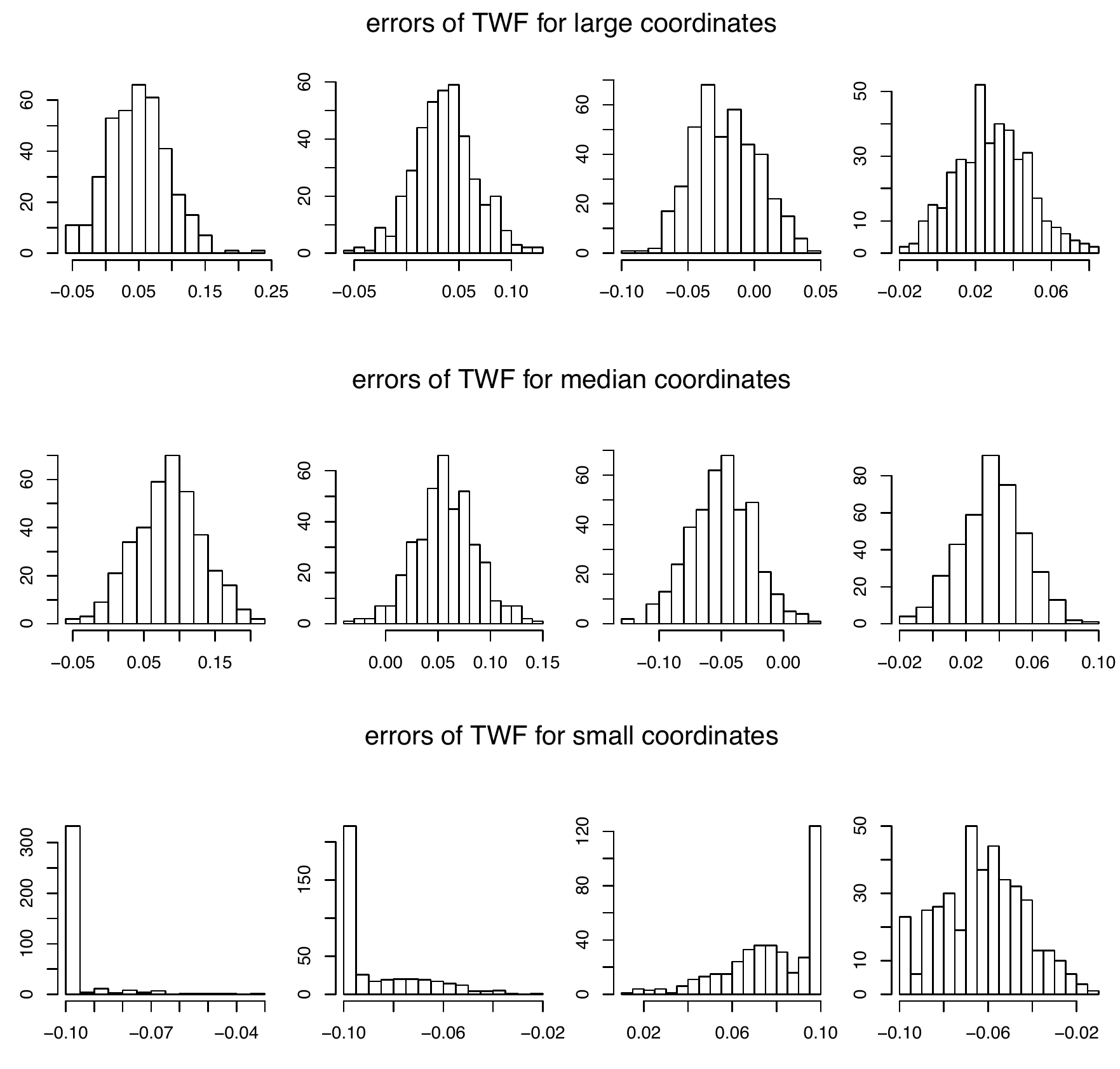}
   \caption{Histograms of TWF errors for large $\beta_k$ (1st row), median $\beta_k$ (2nd row), small $\beta_k$ (3rd row).}
   \label{}
\end{subfigure}\vspace{0.7cm}
\begin{subfigure}[b]{1\textwidth}
   \centering
   \includegraphics[width=0.7\linewidth]{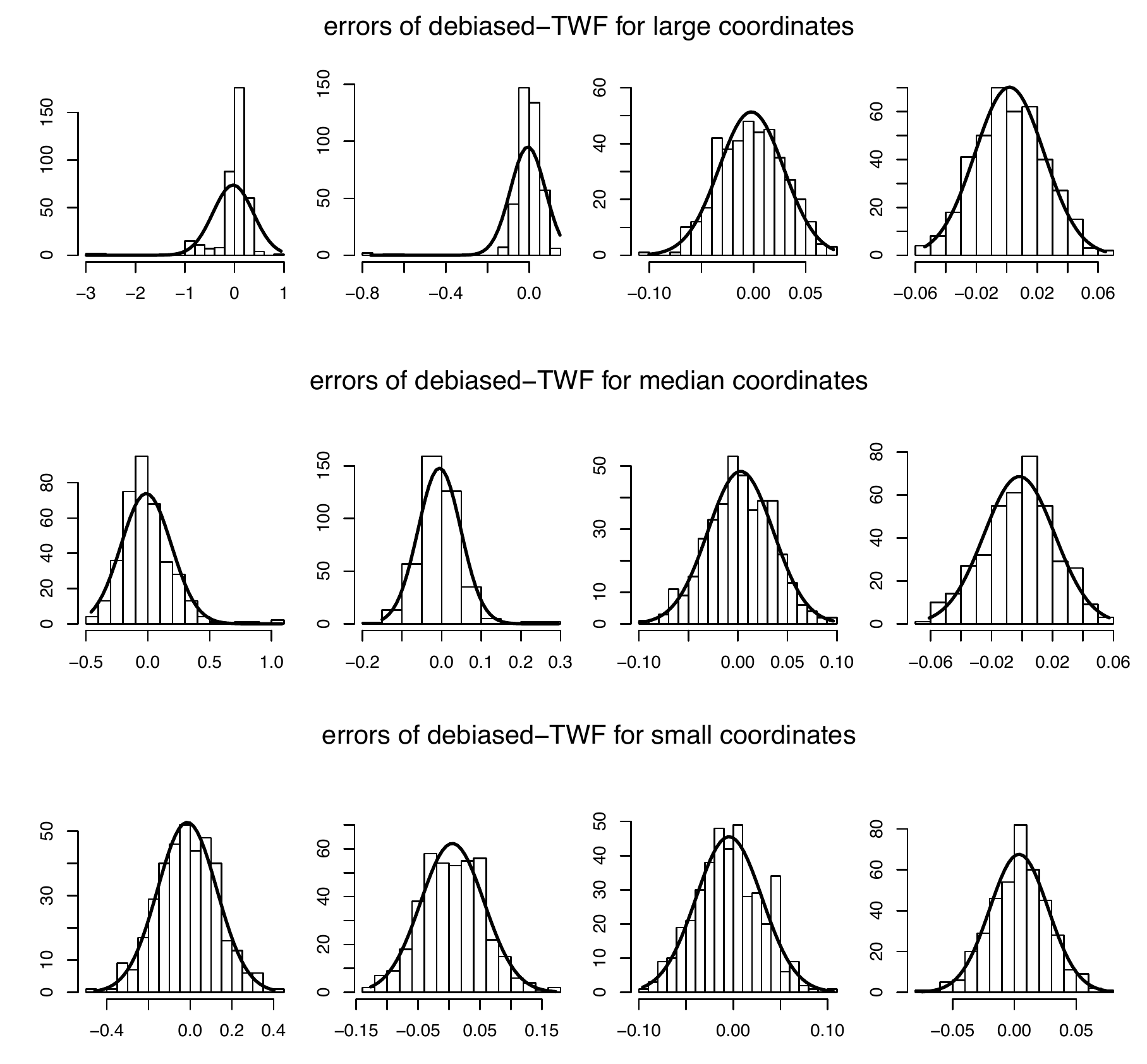}
   \caption{Histograms of debiased TWF errors for large $\beta_k$ (4th row), median $\beta_k$ (5th row), small $\beta_k$ (6th row). }
\end{subfigure}

\caption{ From left to right, plots correspond to simulation settings (n=2000, s=100, NSR=0.3), (n=3000, s=100, NSR=0.3), (n=4000, s=100, NSR=0.3), (n=5000, s=100, NSR=0.3) with p=1000 fixed.}
\end{figure}

\begin{figure}[!htbp]
\centering
\begin{subfigure}[b]{1\textwidth}
   \centering
   \includegraphics[width=0.7\linewidth]{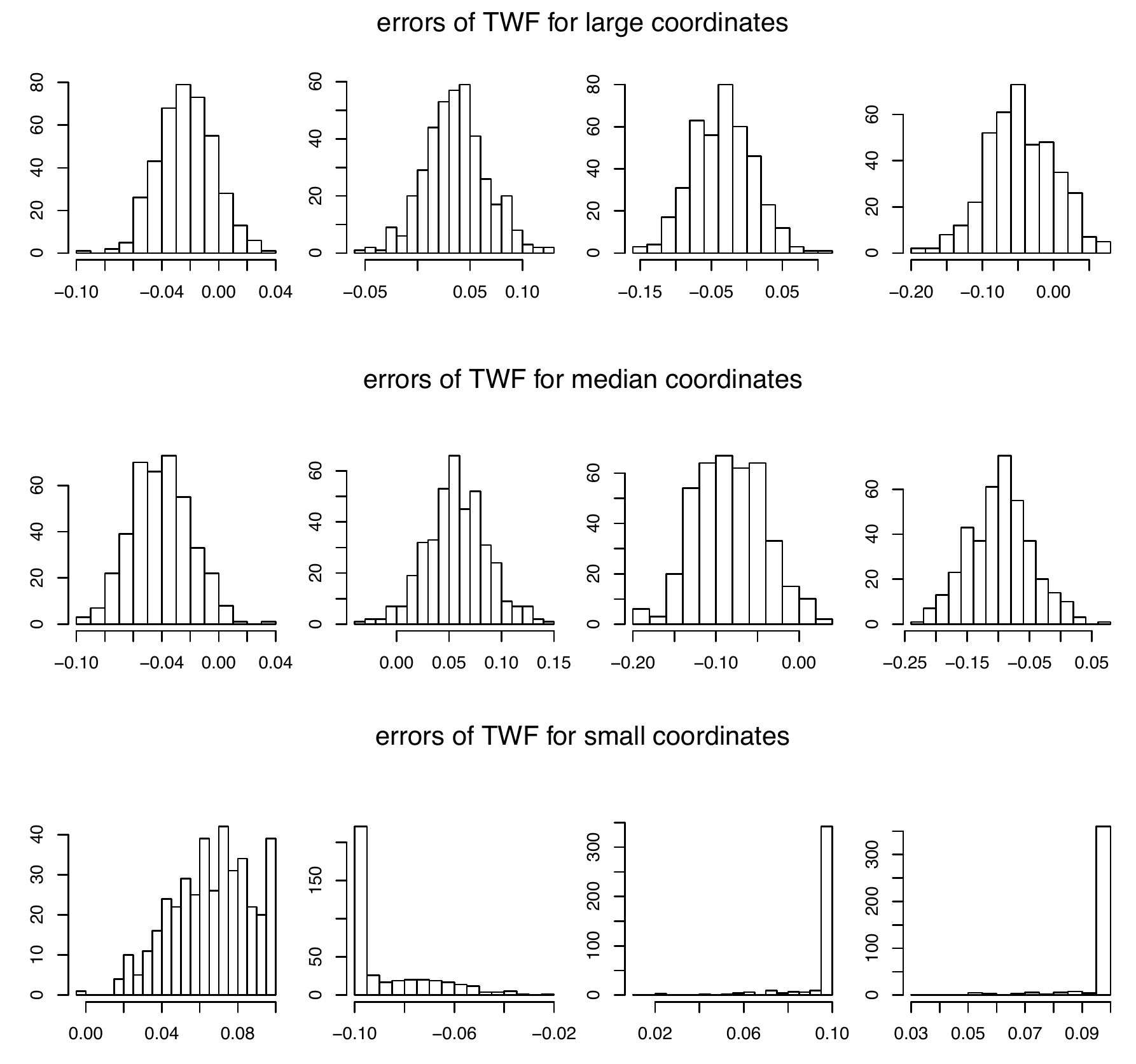}
   \caption{Histograms of TWF errors for large $\beta_k$ (1st row), median $\beta_k$ (2nd row), small $\beta_k$ (3rd row).}
   \label{}
\end{subfigure}
\begin{subfigure}[b]{1\textwidth}
   \centering
   \includegraphics[width=0.7\linewidth]{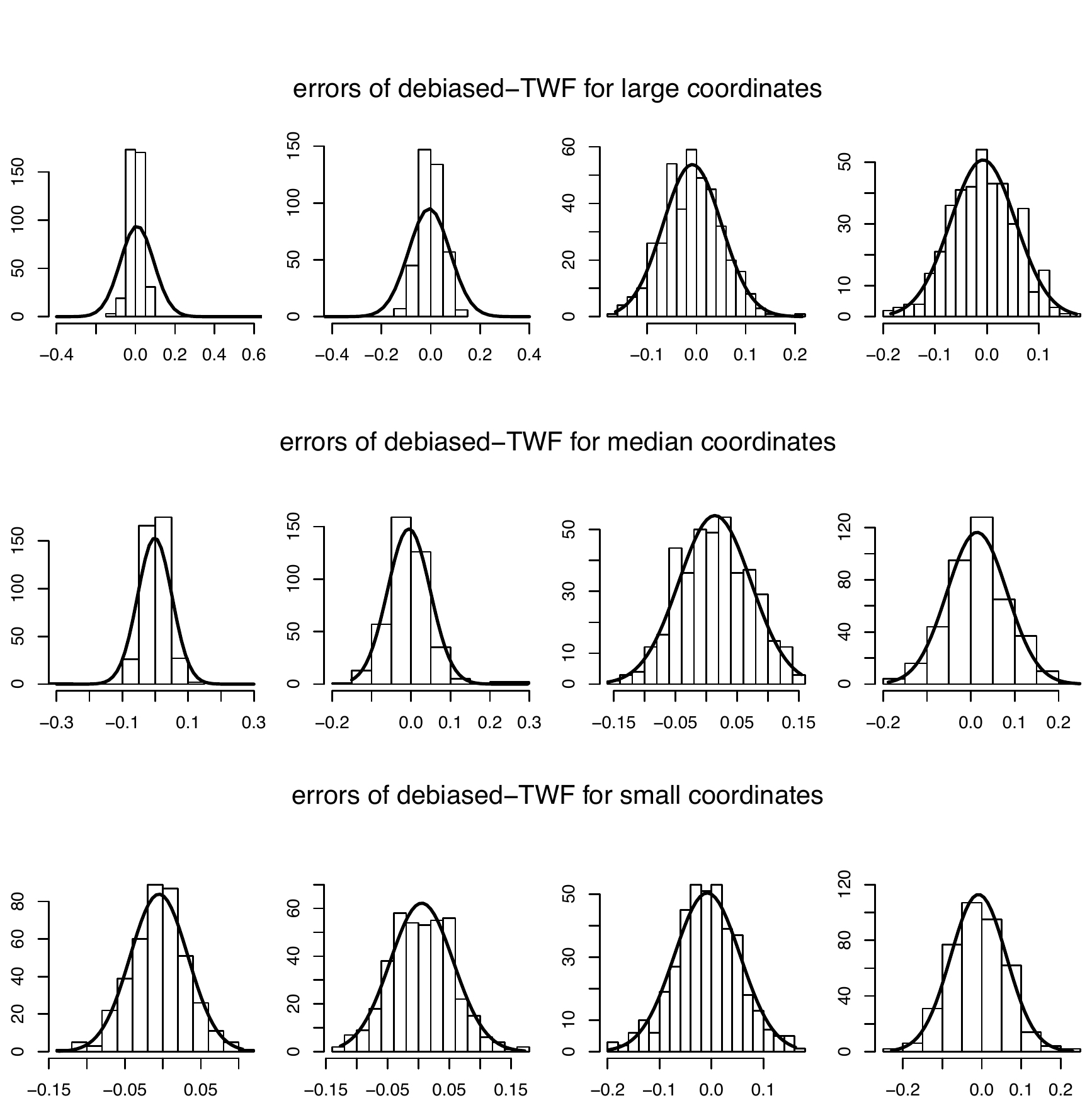}
   \caption{Histograms of debiased TWF errors for large $\beta_k$ (4th row), median $\beta_k$ (5th row), small $\beta_k$ (6th row). }
\end{subfigure}

\caption{ From left to right, plots correspond to simulation settings (n=3000, s=100, NSR=0.2), (n=3000, s=100, NSR=0.3), (n=3000, s=100, NSR=0.4), (n=3000, s=100, NSR=0.5) with p=1000 fixed.}
\end{figure}
\restoregeometry

When simulating to check the accuracy of Theorem \ref{thm-2} ($\bm{ii}$), we use the confidence interval $\bigg(\hat{\beta}^{swap}_k-\frac{\sigma \Phi^{-1}(0.98)}{\sqrt{n}}\sqrt{\frac{\tau_{1k}^2\tau_{2k}^2}{\tau_{1k}^2+\tau_{2k}^2}},\ \hat{\beta}^{swap}_k-\frac{\sigma \Phi^{-1}(0.98)}{\sqrt{n}}\sqrt{\frac{\tau_{1k}^2\tau_{2k}^2}{\tau_{1k}^2+\tau_{2k}^2}}\bigg)$. For each setting, a new $\bbeta$ is generated, and we repeat constructing the confidence interval for 200 times with independently generated $(\X, \bm{\varepsilon}, \y)$. Note, the theoretic coverage probability is $2\times 0.98 -1 - \frac{102}{n}$, in which we count the term $\epsilon_n''$. In asymptotic the term $\epsilon_n''$ is negligible, yet in our simulation $\frac{102}{n} \in (1.36\%, 2\%)$. Therefore, the theoretic coverage probabilities are ranging from 94\% to 94.64\% depending on the variation of n. Table 2 shows that most averagE coverage probabilities are slightly below their theoretical values. The reason for this phenomenon lies in the extra term $\frac{\epsilon_n'}{\sqrt{n}}$ in Theorem \ref{thm-2} $\bm{(ii)}$. Again, in asymptotic $\frac{\epsilon_n'}{\sqrt{n}}$ is negligible compared to $\frac{\sigma r}{\sqrt{n}}\sqrt{\frac{\tau_{1k}^2\tau_{2k}^2}{\tau_{1k}^2+\tau_{2k}^2}}$, but in our simulation these two terms are of the same order. Thus, leaving the term $\frac{\epsilon_n'}{\sqrt{n}}$ out has diminished the coverage probability to certain extend. Despite this flaw, the results in Table 2 imply that Theorem \ref{thm-2} $\bm{(ii)}$ is informative of the actual performance of the debiased TWF. We can see the general trends: the average coverage probabilities get better as n increases or s decreases, and not sensitive to $\sigma$.

\begin{table}[!htbp]
\centering
\begin{tabular}{lll|llll}
\hline
n    & s  & $\sigma$ & all coor & large coor & median coor & small coor \\ \hline
5000 & 40 & 5     & 92.863   & 92.5       & 94.75       & 93.25      \\
6000 & 40 & 5     & 93.2885  & 93.625     & 92.625      & 94.375     \\
6000 & 50 & 5     & 92.7765  & 93         & 92.625      & 94.25      \\
7500 & 50 & 5     & 93.4155  & 93.875     & 92.875      & 93.25      \\
5000 & 40 & 10    & 92.912   & 92.375     & 94.125      & 92.375     \\
6000 & 40 & 10    & 93.664   & 92.75      & 94.75       & 93.25      \\
6000 & 50 & 10    & 92.5935  & 92.25      & 92.25       & 91.625     \\
7500 & 50 & 10    & 93.421   & 93.875     & 93.875      & 95.125     \\ \hline
\end{tabular}
\caption{Mean coverage probabilities of the confidence interval in Theorem 3.2 $\bm{(ii)}$ under various settings. "coor" is short for "coordinate".}
\end{table}

\section{Discussion}
We propose in this work a general approach for drawing statistical inferences on the sparse signal in phase retrieval. A new estimator $\hat{\beta}_k^{swap}$ for the individual signal coordinate $\beta_k$ has been constructed by adding a bias-correction term to the TWF estimator. With mild assumptions on $\X$ and $\bbeta$ and sample size requirement (\ref{assumption2}), $\hat{\beta}_k^{swap}$ has asymptotic Gaussian distribution centered at $\beta_k$. This property allows construction of confidence intervals with approximately preassigned probabilities as well as hypothesis testing on the signal of interest. Our new estimator can also be used as a point estimator. Compared with the plain TWF estimator, $\hat{\beta}_k^{swap}$ achieves asymptotic unbiasness at the cost of a slightly larger variance. 

There remain some open problems. For instance, can we draw statistical inferences on more complicated functions of $\bbeta$, such as a group of coordinates (similar to that in \cite{mitra2016benefit}) or a multidimensional-valued function? Can this method be extended to Fourier designs, which are more applicable? The former is difficult since it involves non-convex optimization over matrices. The later is even more challenging since the Fourier phase retrieval problem is generally considered not solved. And hence we do not have an initial estimator available yet. In summary, there is still long way to go before the Fourier phase retrieval problem is solved and related statistical inferences can be drawn.

\section{Proofs}

{\sc Proof of Theorem \ref{thm-1}.} \\
For the objective function (\ref{objective1}), we have
\begin{equation*}
    \nabla f(\tb)=\frac{1}{n}\sum_{j=1}^n \big[(\bxj^T\tb)^2-y_j\big](\bxj^T\tb)\bxj
\end{equation*}

Let $\Delta=\tb-\bs$. 
\begin{align}
    &\hat{\beta}_k-\beta_k^* = \Tilde{\beta}_k - \beta_k^* + \twk^T \nabla f(\tb) \nonumber\\
    &= \ek^T\Delta + \frac{1}{n}\sum_{j=1}^n \big[(\bxj^T\tb)^2-y_j\big](\bxj^T\tb)(\bxj^T\twk) \nonumber\\
    &= \ek^T\Delta + \frac{1}{n}\sum_{j=1}^n \big[(\bxj^T\tb)^2-(\bxj^T\bs)^2- \varepsilon_j\big](\bxj^T\tb)(\bxj^T\twk) \nonumber\\
    &= \ek^T\Delta +\frac{1}{n}\sum_{j=1}^n(\bxj^T\Delta)^3(\bxj^T\twk) + \frac{3}{n}\sum_{j=1}^n(\bxj^T\Delta)^2(\bxj^T\twk)(\bxj^T\bs) \nonumber\\ 
    &\quad + \frac{2}{n}\sum_{j=1}^n(\bxj^T\Delta)(\bxj^T\twk)(\bxj^T\bs)^2 -\frac{1}{n}\sum_{j=1}^n \varepsilon_j(\bxj^T\tb)(\bxj^T\twk) \nonumber\\
    &= \ek^T\Delta + \frac{2}{n}\sum_{j=1}^n(\bxj^T\Delta)(\bxj^T \wk)(\bxj^T\bs)^2 + \frac{2}{n}\sum_{j=1}^n(\bxj^T\Delta)\big[\bxj^T(\twk-\wk)\big](\bxj^T\bs)^2 \nonumber\\
    &\quad +\frac{1}{n}\sum_{j=1}^n(\bxj^T\Delta)^3(\bxj^T\twk) + \frac{3}{n}\sum_{j=1}^n(\bxj^T\Delta)^2(\bxj^T\twk)(\bxj^T\bs) -\frac{1}{n}\sum_{j=1}^n \varepsilon_j(\bxj^T\tb)(\bxj^T\twk)
    \label{terms}
\end{align}
We will bound these terms separately.\\

Note $supp(\tb) \subseteq S := supp(\bbeta)$ with probability at least $1-\frac{46}{n}-10e^{-s}-\frac{\log n}{np^2}$ when $t\asymp \log\bigg(\frac{\lnorm^2\sqrt{n}}{\sigma\sqrt{s\log p}}\bigg)$. Thus, we have $supp(\Delta)\subseteq S$ with high probability. 
\begin{align*}
    \wk &= -\frac{1}{2}\big(\|\bs\|^2_2 I+2\bs {\bs}^T\big)^{-1}\ek \nonumber\\
    &= -\frac{1}{2\|\bs\|_2^2}\big(I-\frac{2}{3}\frac{\bs}{\|\bs\|_2}\frac{{\bs}^T}{\|\bs\|_2}\big)\ek \nonumber\\
    &= -\frac{1}{2\|\bs\|_2^2}\ek + \frac{\beta_k^*}{3\|\bs\|_2^4}\bs
\end{align*}
Then $supp(\wk)\subseteq S\cup \{k\}$. Similarly, $supp(\twk)\subseteq S\cup \{k\}$. Let $\Bar{S}=S\cup \{k\}$.\\
Checking the supports of $\Delta$, $\wk$ and $\twk$ grants the applicability of Lemma \ref{lemma-A} here, which implies
\begin{align}
    \left\vert \wk^T \Bigg(\frac{1}{n}\sum_{j=1}^n (\bxj^T\bs)^2\bxj\bxj^T - (\|\bs\|^2_2 I+2\bs {\bs}^T)\Bigg) \Delta \right\vert \nonumber\\
    = \left\vert \wk^T \Bigg(\frac{1}{n}\sum_{j=1}^n (\bxj^T\bs)^2{\bxj}_{\Bar{S}}{\bxj}_{\Bar{S}}^T  - (\|\bs\|^2_2 I_{\Bar{S}}+2\bs {\bs}^T)\Bigg) \Delta \right\vert \leq \delta \lnorm^2 \|\Delta\|_2 \|\wk\|_2
\end{align}
with probability at least $1-\frac{1}{n}$, provided $n \geq C(\delta) (s+1)log(s+1)$. Here $C(\delta)$ is constant only depending on $\delta$, and $I_{\Bar{S}}$ is a diagonal matrix with the diagonal elements in $\Bar{S}$ equal to 1, and others equal to 0.

By (\ref{wtrue}) and simple algebra, we have
\begin{equation*}
    \ek^T\Delta+2\wk^T(\|\bs\|^2_2 I_{\Bar{S}}+2\bs {\bs}^T)\Delta=0
\end{equation*}
 
Combining the above two formulae, the first two terms in (\ref{terms}) can be bounded by
\begin{equation}
   \left\vert \ek^T\Delta + \frac{2}{n}\sum_{j=1}^n(\bxj^T\Delta)(\bxj^T \wk)(\bxj^T\bs)^2 \right\vert\leq \delta \lnorm^2 \|\Delta\|_2 \|\wk\|_2
   \label{term12}
\end{equation}
with high probability if $n\geq C(\delta) (s+1)log(s+1)$. 

By similar argument, we have 
\begin{equation*}
    \left\vert \twk^T \Bigg(\frac{1}{n}\sum_{j=1}^n (\bxj^T\Delta)^2\bxj\bxj^T - (\|\Delta\|^2_2 I+2\Delta {\Delta}^T)\Bigg) \Delta \right\vert \leq \delta\|\Delta\|_2^3 \|\twk\|_2 
\end{equation*}
with probability at least $1-\frac{1}{n}$ if $n \geq C(\delta) (s+1)log(s+1)$. Thus, the fourth term in (\ref{terms}) 
is bounded by
\begin{align}
    \left\vert \frac{1}{n}\sum_{j=1}^n(\bxj^T\Delta)^3(\bxj^T\twk) \right\vert &\leq 3\|\Delta\|_2^2 |\Delta^T\twk| + \delta\|\Delta\|_2^3 \|\twk\|_2 \nonumber\\
    &\leq 3\dnorm^3 \|\twk\|_2 + \delta\|\Delta\|_2^3 \|\twk\|_2
    \label{term4}
\end{align}
with high probability. 

Again via similar reasoning, the fifth term in (\ref{terms}) falls within $3\|\Delta\|_2^2 (\twk^T\bs) + 6(\Delta^T\twk)(\Delta^T\bs) \pm 3\delta \|\Delta\|_2^2 \|\twk\|_2 \|\bs\|_2$ with probability at least $1-\frac{1}{n}$ provided $n \geq C(\delta) (s+1)log(s+1)$. And we bound it by
\begin{align}
    \left\vert \frac{3}{n}\sum_{j=1}^n(\bxj^T\Delta)^2(\bxj^T\twk)(\bxj^T\bs) \right\vert &\leq 3\dnorm^2 |\twk^T\bs| + 6|\Delta^T\twk||\Delta^T\bs| + 3\delta \dnorm^2 \|\twk\|_2 \|\bs\|_2 \nonumber\\
    &\leq 9\dnorm^2\|\twk\|_2\|\bs\|_2 + 3\delta \dnorm^2 \|\twk\|_2 \|\bs\|_2
    \label{term5}
\end{align}
The derivations of (\ref{term12})-(\ref{term5}) require a common condition $n \geq C(\delta) (s+1)log(s+1)$ with exactly the same $C(\delta)$.\\

Next, we deal with the third term in (\ref{terms}). 
\begin{align*}
    \frac{1}{n}\sum_{j=1}^n(\bxj^T\Delta)\big[\bxj^T(\twk-\wk)\big](\bxj^T\bs)^2 &= \Delta^T \Bigg(\frac{1}{n}\sum_{j=1}^n (\bxj^T\bs)^2\bxj\bxj^T - (\|\bs\|^2_2 I+2\bs {\bs}^T)\Bigg) (\twk-\wk) \nonumber\\
    &\quad + \lnorm^2\Delta^T(\twk-\wk) + 2(\bbeta^{*T}\Delta)\bigg(\bbeta^{*T}(\twk-\wk)\bigg)
\end{align*}
Applying Lemma \ref{lemma-A} one more time, the third term falls within 
\begin{equation*}
    \lnorm^2\Delta^T(\twk-\wk) + 2(\bbeta^{*T}\Delta)\bigg(\bbeta^{*T}(\twk-\wk)\bigg) \pm \delta \lnorm^2 \|\Delta\|_2 \|\twk-\wk\|_2 
\end{equation*}
with probability at least $1-\frac{1}{n}$, given $n \geq C(\delta) (s+1)log(s+1)$.
The magnitude of $\|\twk-\wk\|_2$ in terms of $\lnorm$ and $\dnorm$ can be estimated by
\begin{align*}
    \twk-\wk = \frac{1}{2} \Bigg(\frac{1}{\lnorm^2}-\frac{1}{\tlnorm^2}\Bigg)\ek + \frac{\Tilde{\beta}_k}{3\tlnorm^4}\Delta + \frac{1}{3}\Bigg(\frac{\Tilde{\beta}_k}{\tlnorm^4}-\frac{\beta^*_k}{\lnorm^4}\Bigg)\bs
\end{align*}
\begin{align}
    \|\twk-\wk\|_2 &\leq \frac{1}{2}\left\vert \frac{1}{\lnorm^2}-\frac{1}{\tlnorm^2}\right\vert + \frac{|\Tilde{\beta}_k|}{3\tlnorm^4}\|\Delta\|_2 + \frac{1}{3}\left\vert \frac{\Tilde{\beta}_k}{\tlnorm^4}-\frac{\beta^*_k}{\lnorm^4}\right\vert \lnorm \nonumber\\
    &= O(\frac{\|\Delta\|_2}{\lnorm^3})
    \label{wtrue-west}
\end{align}
The last equality is because
\begin{align*}
    \left\vert \frac{1}{\lnorm^2}-\frac{1}{\tlnorm^2} \right\vert &= \frac{\|\bs+\Delta\|_2^2-\lnorm^2}{\lnorm^2\tlnorm^2} \leq \frac{2\lnorm\|\Delta\|_2+\|\Delta\|_2^2}{\lnorm^2\tlnorm^2} = O(\frac{\|\Delta\|_2}{\lnorm^3}),
\end{align*}
and
\begin{align*}
    \left\vert \frac{\Tilde{\beta}_k}{\tlnorm^4}-\frac{\beta^*_k}{\lnorm^4} \right\vert &= \frac{1}{\lnorm^4\tlnorm^4}\left\vert \lnorm^4\Tilde{\beta}_k - \tlnorm^4\beta^*_k\right\vert\\
    & \leq \frac{\lnorm^4|\Tilde{\beta}_k-\beta^*_k|}{\lnorm^4\tlnorm^4} + \frac{\left\vert \tlnorm^4-\lnorm^4\right\vert |\beta^*_k|}{\lnorm^4\tlnorm^4} \\
    & \leq \frac{\|\Delta\|_2}{\tlnorm^4} + \frac{\bigg(\lnorm+\|\Delta\|_2\bigg)^4-\lnorm^4}{\lnorm^3\tlnorm^4}\\
    & \leq O(\frac{\|\Delta\|_2}{\lnorm^3}).
\end{align*}
Therefore, we bound the third term in (\ref{terms}) by
\begin{align}
    \left\vert \frac{2}{n}\sum_{j=1}^n(\bxj^T\Delta)\big[\bxj^T(\twk-\wk)\big](\bxj^T\bs)^2 \right\vert &\leq 3\lnorm^2\dnorm\|\twk-\wk\|_2 + \delta\lnorm^2\dnorm\|\twk-\wk\|_2 \nonumber\\
    &= 3\lnorm^2\cdot O(\frac{\dnorm^2}{\lnorm^3}) + \delta\lnorm^2\cdot O(\frac{\dnorm^2}{\lnorm^3})
    \label{term3}
\end{align}
$\|\twk\|_2$ appears in the mean terms of (\ref{term4}), (\ref{term5}) and needs to be bounded. By (\ref{wtrue}) and (\ref{wtrue-west}), 
\begin{align*}
     \|\wk\|_2 &\leq \frac{1}{2\lnorm^2} + \frac{\beta^*_k}{3\lnorm^3} \leq \frac{1}{2\lnorm^2} + \frac{1}{3\lnorm^2} \nonumber\\
     & \leq \frac{5}{6\lnorm^2}
\end{align*}
\begin{align*}
    \|\twk\|_2 &\leq \|\wk\|_2 + O(\frac{\dnorm}{\lnorm^3}) \nonumber\\
    &\leq \frac{5}{6\lnorm^2} + O(\frac{\dnorm}{\lnorm^3})
\end{align*}

So far, the terms that differ $\hat{\beta}_k -\beta^*_k$ from $-\frac{1}{n}\sum_{j=1}^n \varepsilon_j(\bxj^T\tb)(\bxj^T\twk)$ (an asymptotically normal random variable) have been concentrated around their means, with the concentration errors associated with $\delta$. We can set $\delta=\frac{1}{p^2}$, so that all concentration errors vanish in an order faster than $\frac{\sqrt{n}}{p^2}$. And we can just ignore these terms. The only thing left to check is whether these mean terms are negligible after multiplying by $\sqrt{n}$, \textit{i.e.} to show that $\sqrt{n}(\hat{\beta}_k-\beta_k^*)$ is approximately normal as $n\rightarrow\infty$. 
The goal is to show 
\begin{equation*}
    \left\vert \sqrt{n}(\hat{\beta}_k-\beta_k^*) - Z_k \right\vert = o_p(1)
\end{equation*}

By (\ref{terms}), (\ref{term12})-(\ref{term5}), (\ref{term3}), (\ref{Zk}) we have with probability at least $1-\frac{4}{n}$,
\begin{align}
    &\left\vert \sqrt{n}(\hat{\beta}_k-\beta_k^*) - Z_k \right\vert \nonumber\\
    &\leq \sqrt{n}\Bigg( 3\dnorm^3\|\twk\|_2 + 9\dnorm^2\|\twk\|_2\lnorm + 3\lnorm^2\cdot O(\frac{\dnorm^2}{\lnorm^3}) \Bigg) + O(\frac{\sqrt{n}}{p^2}) \nonumber\\
    &\leq \sqrt{n}\Bigg(\frac{5}{2}\frac{\dnorm^3}{\lnorm^2}+ O(\frac{\dnorm^4}{\lnorm^3}) + \frac{15}{2}\frac{\dnorm^2}{\lnorm}+ O(\frac{\dnorm^3}{\lnorm^2}) + O(\frac{\dnorm^2}{\lnorm}) \Bigg) + O(\frac{\sqrt{n}}{p^2})
\end{align}
provided $n \geq C(\frac{1}{p^2}) (s+1)log(s+1)$, which is satisfied under assumption (\ref{assumption1}) and (\ref{assumption2}). Here $C(\frac{1}{p^2})$ is a constant only depending on $\frac{1}{p^2}$.
Together with (\ref{TWF1}) and (\ref{TWF2}) that have been established in \cite{cai2016optimal}, we obtain
\begin{align}
     &\left\vert \sqrt{n}(\hat{\beta}_k-\beta_k^*) - Z_k \right\vert \nonumber\\
     &\leq C_1\frac{\sigma^3}{\lnorm^5}\frac{s\log p\sqrt{s\log p}}{n}+ C_2\frac{\sigma^4}{\lnorm^7}\frac{s^2(\log p)^2}{n\sqrt{n}}+ C_3\frac{\sigma^2}{\lnorm^3}\frac{s\log p}{\sqrt{n}} + C_4\frac{\sqrt{n}}{p^2}
     \label{difference}
\end{align}
with probability at least $1-\frac{50}{n}-10e^{-s}-\frac{M\log n}{np^2}$ for some absolute constants $M, C_1, C_2, C_3, C_4$.\\
Plugging in $\lnorm=O(\sqrt{s})$, the right hand side of (\ref{difference}) becomes $C'_1\frac{(\log p)^{\frac{3}{2}}}{sn} + C'_2\frac{(\log p)^2}{(sn)^{\frac{3}{2}}} + C'_3\frac{\log p}{(sn)^\frac{1}{2}} + C'_4\frac{n^{\frac{1}{2}}}{p^2}$, all of which vanish as $n\rightarrow\infty$ under assumption (\ref{assumption2}). And the asymptotic normality of $\sqrt{n}(\hat{\beta}_k-\beta_k^*)$ is established. $\hfill\square$

\bigbreak
\bigbreak
{\sc Proof of Theorem \ref{thm-2}.} \\
$\bm{(i)}\quad$
It is intuitive that a reasonable estimator would combine $\hat{\beta}_{1k}$ and $\hat{\beta}_{2k}$ so as to integrate both pieces of information. We know that $Z_{1k}$ and $Z_{2k}$ have asymptotic variances $\sigma^2\tau_{1k}^2$ and $\sigma^2\tau_{2k}^2$, respectively. By Theorem \ref{thm-1} and Remark 3.1, 
\begin{align*}
    \hat{\beta}_{1k} := \beta_k^* + \frac{Z_{1k}}{\sqrt{n}} + op(\frac{\sigma\tau_{1k}}{\sqrt{n}}) = \beta_k^* + N(0,\frac{\sigma^2\tau_{1k}^2}{n}) + op(\frac{\sigma\tau_{1k}}{\sqrt{n}})\\
    \hat{\beta}_{2k} := \beta_k^* + \frac{Z_{2k}}{\sqrt{n}} + op(\frac{\sigma\tau_{2k}}{\sqrt{n}}) = \beta_k^* + N(0,\frac{\sigma^2\tau_{2k}^2}{n}) + op(\frac{\sigma\tau_{2k}}{\sqrt{n}})
\end{align*}
Judging from these two formulae, any convex combination of $\hat{\beta}_{1k}$ and $\hat{\beta}_{2k}$ remains asymptotically unbiased (has asymptotic mean equal to $\beta^*_k$) and possibly attain a smaller asymptotic variance. Suppose we have the final estimator given by 
\begin{equation*}
    \hat{\beta}_k^{swap} = a\hat{\beta}_{1k} + (1-a)\hat{\beta}_{2k}\quad for\  a\in (0,1).
\end{equation*}
then, 
\begin{align*}
    \sqrt{n}(\hat{\beta}_k^{swap}-\beta_k^*) = a\cdot Z_{1k} + (1-a)\cdot Z_{2k} + op(1) 
\end{align*}

It is easy to verify that $COV(Z_{1k}, Z_{2k})=0$, and hence the asymptotic variance of $\hat{\beta}_k^{swap}$ is $\sigma^2\bigg( a^2\tau_{1k}^2 + (1-a)^2\tau_{2k}^2 \bigg)$, which is a quadratic form in $a$. When $a=\frac{\tau_{2k}^2}{\tau_{1k}^2+\tau_{2k}^2}$, the asymptotic variance attains minimum, in which case $\hat{\beta}_k^{swap}$ has approximately asymptotic distribution $N(0,\sigma^2\frac{\tau_{1k}^2\tau_{2k}^2}{\tau_{1k}^2+\tau_{2k}^2})$.

\bigbreak
$\bm{(ii)}\quad$
Let $a=\frac{\tau_{2k}^2}{\tau_{1k}^2+\tau_{2k}^2}$. By Lemma \ref{lemma-A}, for $n$ satisfying (\ref{assumption2}),
\begin{align*}
     \mathbb{P}\Bigg\{a^2\left\lvert \frac{Var(Z_{1k})}{\sigma^2}-\tau_{1k}^2 \right\rvert \leq \frac{a^2}{p^2}\|\tb_1\|_2^2\|\ftwk\|_2^2\Bigg\} \geq 1-\frac{1}{n}\\
     \mathbb{P}\Bigg\{(1-a)^2\left\lvert \frac{Var(Z_{2k})}{\sigma^2}-\tau_{2k}^2 \right\rvert \leq \frac{(1-a)^2}{p^2}\|\tb_2\|_2^2\|\stwk\|_2^2\Bigg\} \geq 1-\frac{1}{n}
\end{align*}
Let $Z_k^{swap} = a\cdot Z_{1k} + (1-a)\cdot Z_{2k}$, then
\begin{equation}
     \mathbb{P}\Bigg\{\left\lvert \frac{Var(Z_k^{swap})}{\sigma^2}-a^2\tau_{1k}^2-(1-a)^2\tau_{2k}^2 \right\rvert \leq \frac{1}{p^2}\bigg(a^2\|\tb_1\|_2^2\|\ftwk\|_2^2+(1-a)^2\|\tb_2\|_2^2\|\stwk\|_2^2\bigg)\Bigg\} \geq 1-\frac{2}{n}
     \label{prf_thm2_1}
\end{equation}
Similar to the argument in Theorem \ref{thm-1},  
\begin{equation}
    \|\tb_1\|_2^2\|\ftwk\|_2^2 \leq \bigg(\lnorm+2\dnorm\bigg)^2\bigg(\frac{5}{6\lnorm^2}+O(\frac{\dnorm}{\lnorm^3})\bigg)^2 \asymp \frac{25}{36 s} + O(\sqrt{\frac{\log p}{n s^3}})
    \label{prf_thm2_2}
\end{equation}
Plugging (\ref{prf_thm2_2}) into (\ref{prf_thm2_1}), together with the fact $a^2+(1-a)^2\leq \frac{1}{2}$, we have 
\begin{align*}
     \mathbb{P}\Bigg\{\left\lvert \frac{Var(Z_k^{swap})}{\sigma^2}-a^2\tau_{1k}^2-(1-a)^2\tau_{2k}^2 \right\rvert \leq \frac{1}{p^2}\bigg(\frac{25}{72 s}+ O(\sqrt{\frac{\log p}{n s^3}})\bigg)\Bigg\} \geq 1-\frac{2}{n}.
\end{align*}
Further plugging the value of $a$, 
\begin{align*}
     \mathbb{P}\Bigg\{ Var(Z_k^{swap}) \leq \sigma^2 \frac{\tau_{1k}^2\tau_{2k}^2}{\tau_{1k}^2+\tau_{2k}^2} + O(\frac{1}{p^2})\Bigg\} \geq 1-\frac{2}{n}.
\end{align*}
By (\ref{beta_swap}), we get
\begin{align*}
     \mathbb{P}\Bigg\{\left\lvert \hat{\beta}^{swap}_k-\beta_k^*\right\rvert \leq \frac{1}{\sqrt{n}}|Z_k^{swap}|+ \frac{\epsilon_n'}{\sqrt{n}}\Bigg\} \geq 1-2\epsilon_n''
\end{align*}
Let $\Phi(\cdot)$ be the cumulative distribution function of standard normal distribution, for $\forall r > 0$,
\begin{align*}
     \mathbb{P}\Bigg\{ \frac{1}{\sqrt{n}}|Z_k^{swap}| \leq \frac{r\sqrt{Var(Z_k^{swap})}}{\sqrt{n}}\Bigg\} = 2\Phi(r)-1
\end{align*}
The above three formulae imply (\ref{thm2_2}), which means the interval $\bigg(\hat{\beta}^{swap}_k-\frac{\sigma r}{\sqrt{n}}\sqrt{\frac{\tau_{1k}^2\tau_{2k}^2}{\tau_{1k}^2+\tau_{2k}^2}}+ \frac{\epsilon_n'}{\sqrt{n}},\ \hat{\beta}^{swap}_k-\frac{\sigma r}{\sqrt{n}}\sqrt{\frac{\tau_{1k}^2\tau_{2k}^2}{\tau_{1k}^2+\tau_{2k}^2}}+ \frac{\epsilon_n'}{\sqrt{n}}\bigg)$ has asymptotic coverage probability at least $2\Phi(r)-1$. 
\bigbreak

$\bm{(iii)}\quad$
By Remark 3.1, $\frac{\tau_{1k}^2\tau_{2k}^2}{\tau_{1k}^2+\tau_{2k}^2} \leq \frac{1}{2}\max\{\tau_{1k}^2, \tau_{2k}^2\}\asymp \frac{3}{8s} + O(\sqrt{\frac{\log p}{n s^3}})$ for all $k=1,2,\cdots,p$. Plus (\ref{thm2_2}) holds uniformly over k, we obtain 
\begin{equation*}
    \liminf\limits_{n\rightarrow\infty}\mathbb{P}\Bigg\{\max_{k\in [p]}\left\lvert \sqrt{n}(\hat{\beta}^{swap}_k-\beta_k^*)\right\rvert \leq \sqrt{\frac{3}{8s}}\sigma r\Bigg\} \geq 2\phi(r)-1 
\end{equation*}
and then replace r by $\Phi(1-\frac{\alpha}{2})$.\\
It is easy to verify that the asymptotic distribution of $\hat{\bbeta}^{swap}$ is multinormal $N(\bs, \frac{\sigma^2}{n}\bm{V})$, where 
\begin{align*}
    V_{kl} &= \frac{1}{\sigma^2} Cov(Z_k^{swap}, \Tilde{Z}_l) = \frac{1}{\sigma^2} Cov\bigg(a_k Z_{1k} + (1-a_k)Z_{2k}, a_l Z_{1l} + (1-a_l)Z_{2l}\bigg)\\
    &= \frac{1}{\sigma^2}\bigg[a_k a_l Cov(Z_{1k}, Z_{1l}) + (1-a_k)(1-a_l) Cov(Z_{2k},Z_{2l})\bigg]\\
    &= a_k a_l \bigg[\|\tb_1\|_2^2(\ftwk^T\ftwl)+ 2(\tb_1^T\ftwk)(\tb_1^T\ftwl)\bigg] + (1-a_k)(1-a_l) \bigg[\|\tb_2\|_2^2(\stwk^T\stwl)+ 2(\tb_2^T\stwk)(\tb_2^T\stwl)\bigg]
\end{align*}
By linear algebra, 
\begin{align*}
    \sup_{\bm{h}\neq 0, \bm{h}\in \mathbb{R}^p} \frac{\left\lvert \bm{h}^T(\hat{\bbeta}^{swap}-\bs)\right\rvert ^2}{\bm{h}^T\bm{V}\bm{h}} = (\hat{\bbeta}^{swap}-\bs)^T \bm{V}^{-1} (\hat{\bbeta}^{swap}-\bs) \xrightarrow{\text{d}} \frac{\sigma^2}{n} \chi_p^2
\end{align*}
Therefore, the coverage probability in the worst direction is bounded below,
\begin{equation*}
    \liminf\limits_{n\rightarrow\infty}\mathbb{P}\Bigg\{ \sup_{\bm{h}\neq 0, \bm{h}\in \mathbb{R}^p} \frac{\left\lvert \bm{h}^T(\hat{\bbeta}^{swap}-\bs)\right\rvert ^2}{\bm{h}^T(\frac{\sigma^2\bm{V}}{n})\bm{h}} \leq \chi^2_{p,\alpha} \Bigg\} \geq 1-\alpha 
\end{equation*}
It further implies, for $\forall h\neq 0$, $h\in \mathbb{R}^p$,
\begin{equation*}
    \liminf\limits_{n\rightarrow\infty}\mathbb{P}\Bigg\{  \left\lvert \bm{h}^T(\hat{\bbeta}^{swap}-\bs)\right\rvert \leq \sqrt{\frac{\sigma^2}{n}\chi^2_{p,\alpha} {\bm{h}^T \bm{V} \bm{h}}}\Bigg\} \geq 1-\alpha 
\end{equation*}
$\hfill\square$

\section*{Acknowledgements}
The authors would like to thank Cun-Hui Zhang for several constructive advises and enlightening discussions. The authors also would like to thank Pierre Bellec for the enlightening discussions. 

\clearpage
\section*{Appendix}
\renewcommand{\thesection}{A}

The theorem and lemma below are stated and proved in \cite{cai2016optimal}.

\begin{theorem}\label{thm-A} \cite{cai2016optimal}
Suppose the tuning parameters in the thresholded Wirtinger algorithm are suitably chosen, and the sample size $n\geq K(1+\frac{\sigma}{\|\beta\|_2^2})^2 s^2\log(np)$ for some absolute constant $K > 0$, Let $S=support(\bbeta)$, then
\begin{align*}
    \inf_{\|\bbeta\|_0=s}\mathbb{P}_{(\X,\y|\bbeta)}\Bigg\{supp(\tb^{(t)})\subseteq S\ and\ \min_{i=0,1}\|\tb^{(t)}-(-1)^{i}\bbeta\|_2\leq\frac{1}{6}(1-\frac{w}{16})^t\lnorm+C\frac{\sigma}{\lnorm}\sqrt{\frac{s\log{p}}{n}}\Bigg\}\\
    > 1-\frac{46}{n}-10e^{-s}-\frac{t}{np^2}
\end{align*}
for some absolute constant $C > 0$, where $w$ is the gradient descent step size. 
When $\frac{\sigma}{\lnorm^2}=o(\sqrt{\frac{n}{\log n}})$ and is unknown, we can estimate $\lnorm^2$ by
\begin{align*}
    \phi^2 := \widehat{\lnorm^2}=\frac{1}{n}\sum_{j=0}^{n}y_j
\end{align*}
and define 
\begin{align*}
    \hat{\sigma}=\sqrt{(\frac{1}{n}\sum_{j=0}^{n}y_j^2)-3\phi^4}.
\end{align*}
Then with probability at least $1-\frac{1}{n}$, there holds $\frac{\hat{\sigma}}{\phi^2}\asymp \frac{\sigma}{\lnorm^2}$. If the sample size $n\geq K(1+\frac{\hat{\sigma}}{\phi^2})^2 s^2\log(np)$, the the above claim holds with the first term on the right hand side $\frac{46}{n}$ replaced by $\frac{47}{n}$. \end{theorem}

\bigbreak

\begin{lemma}\label{lemma-A} \cite{cai2016optimal}
Suppose $\bxj$ are \textit{i.i.d.} $N(0, I_{p\times p})$. Then on an event with probability at least $1-\frac{1}{n}$, we have 
\begin{equation*}
    \left\lVert \frac{1}{n}\sum_{j=1}^n (\bxj^T\bb)^2{\bxj}_{\Bar{S}}{\bxj}_{\Bar{S}}^T - (\|\bb\|^2_2 I_{\Bar{S}}+2\bb {\bb}^T) \right\rVert \leq \delta \|\bb\|_2^2
\end{equation*}
provided $n\geq C(\delta)s\log s$, where $C(\delta)$ is constant only depending on $\delta$. Here, $I_S$ is a diagonal matrix with the diagonal elements in $S$ equal to 1, whereas others equal to 0. And $supp(\bb)\subset S$.
\end{lemma}

\clearpage
\printbibliography

\end{document}